\documentclass[aps,prb,twocolumn,floatfix,footinbib,showpacs,superscriptaddress]{revtex4-1}
\usepackage{graphicx}
\usepackage{amsfonts,amsmath,amssymb}
\usepackage{amsthm}
\usepackage{dsfont,bm}
\usepackage{color}
\usepackage{soul} 
\usepackage{amsbsy}
\usepackage[colorlinks=true,linkcolor=blue,pagecolor=blue,filecolor=blue,menucolor=blue,urlcolor=blue,citecolor=blue,anchorcolor=blue]{hyperref}%
\usepackage{multirow}
\usepackage{mathrsfs}
\usepackage{txfonts}
\usepackage{amsbsy}
\usepackage{times} 
\usepackage{dcolumn}

\newcommand{\ua}{\uparrow}
\newcommand{\da}{\downarrow}

\usepackage{lineno}

\begin{document}

\title{Symmetry of superconducting correlations in displaced bilayers of graphene}

\author{Mohammad Alidoust}
\affiliation{Department of Physics, K.N. Toosi University of Technology, Tehran 15875-4416, Iran}
\author{Morten Willatzen}
\affiliation{Beijing Institute of Nanoenergy and Nanosystems, Chinese Academy of Sciences,
No. 30 Xueyuan Road, Haidian District, Beijing 100083, China}
\affiliation{Department of Photonics Engineering, Technical University of Denmark, Kongens Lyngby, DK-2800, Denmark}
\author{Antti-Pekka Jauho}
\affiliation{Center for Nanostructured Graphene (CNG), DTU Nanotech, Technical University of Denmark, DK-2800 Kongens Lyngby, Denmark}

\begin{abstract}
Using a Green's function approach, we study phonon-mediated superconducting pairing symmetries that may arise in bilayer graphene where the monolayers are displaced in-plane with respect to each other. We consider a generic coupling potential between the displaced graphene monolayers, which is applicable to both shifted and commensurate twisted graphene layers; study intralayer and interlayer phonon-mediated BCS pairings; and investigate AA and AB(AC) stacking orders. Our findings demonstrate that at the charge neutrality point, the dominant pairings in both AA and AB stackings with intralayer and interlayer electron-electron couplings can have even-parity $s$-wave class and odd-parity $p$-wave class of symmetries with the possibility of invoking equal-pseudospin and odd-frequency pair correlations. At a finite doping, however, the AB (and equivalently AC) stacking can develop pseudospin-singlet and pseudospin-triplet $d$-wave symmetry, in addition to $s$-wave, $p$-wave, $f$-wave, and their combinations, while the AA stacking order, similar to the undoped case, is unable to host the $d$-wave symmetry. When we introduce a generic coupling potential, applicable to commensurate twisted and shifted bilayers of graphene, $d$-wave symmetry can also appear at the charge neutrality point. Inspired by a recent experiment where two phonon modes were observed in a twisted bilayer graphene, we also discuss the possibility of the existence of two-gap superconductivity, where the intralayer and interlayer phonon-mediated BCS picture is responsible for superconductivity. These analyses may provide a useful tool in determining the superconducting pairing symmetries and mechanism in bilayer graphene systems.
\end{abstract}

\date{\today}

\maketitle

\section{introduction}

Graphene, a two-dimensional honeycomb arrangement of covalently bound carbon atoms, can be used as a basic building block for novel multilayer systems coupled by interlayer van der Waals forces. These weak interlayer bonding forces allow the possibility to create different ordering stacks that possess dramatically different low-energy properties, offering low-cost tools for rich band structure and material engineering perspectives. The ordering of the stacks can be controlled externally by growth techniques or by the application of tensile stress to laterally shift and/or rotate the layers in plane with respect to each other \cite{Butz, Kisslinger,Hass,Borysiuk, Liu}.

Depending on the stacking order, the multilayers can have a number of different configurations. Bilayer graphene (BLG) where the top and bottom layers are mutually rotated is an example of introducing additional features by a simple geometric action. It has been shown that at low angles of the twist, the BLG develops a highly nontrivial band structure \cite{tbg-2,tbg-1,tbg0,tbg1,tbg2,tbg3,tbg4,tbg5,tbg6,tbg8,tbg9,tbg10,tbg11,tbg12,tbg13,tbg14,2ph-el_exp}. An efficient theory tool to approach such a problem and describe its underlying physics is the effective Hamiltonian method. This technique assumes that the two pristine graphene layers are unaffected under the displacement and all changes can be translated into a new coupling potential between the two layers \cite{ray1,ray2,ray_slip}.

Physically, one can expect that the charge carriers in BLG experience different interactions due to different phonon modes generated within or between the layers. The two pristine graphenes have identical phonon modes (due to the intralayer covalent bonds) while a different mode appears due to the interlayer coupling (van der Waals bonds). Note that our conclusions depend on having two different coupling strengths (intralayer and interlayer couplings), and the microscopic details of the coupling do not matter.  For example, the interlayer coupling could have a significant covalent component. \cite{tbg2,tbg3,tbg4,tbg5,tbg6,tbg8,tbg9,tbg10,tbg11,tbg12,tbg13,tbg14,2ph-el_exp}. 
In particular, recent experiments demonstrated the different intralayer and interlayer electron-phonon interactions by using a Raman spectroscopy technique: two resonance peaks in the Raman modes  were observed by tuning the energy of excitation with IR and UV photons \cite{2ph-el_exp}.

The electron-phonon coupling affects, for instance, the electronic mobility, thermal conductivity, and the (possible) superconducting phases. A recent landmark experiment showed intrinsic superconductivity with a critical temperature of $\sim 1.7$~K in a twisted BLG, away from the charge neutrality point $\mu>0$ at a very low carrier density of $10^{11}$~cm$^{-2}$. Interestingly, the general features of the phase diagram seemed to share the same phenomenology as high $T_c$ superconductors \cite{herrero}. Soon after, another experiment \cite{C.Dean_exp} (and recently few others \cite{grexp1,grexp2,grexp3}) and a very large number of theoretical papers have appeared, many of which focus  on the phonon-mediated single-gap superconductivity, and addressed various aspects of the observed superconducting phase (see, e.g. Refs. [\onlinecite{teor_tbg1,teor_tbg2,teor_tbg3,teor_tbg4,teor_tbg5,teor_tbg7,
teor_tbg9,teor_tbg11,teor_tbg12,teor_tbg13,teor_tbg14,teor_tbg15,teor_tbg16,teor_tbg17,teor_tbg18,teor_tbg19,teor_tbg20,teor_tbg21,YWChoi}]). However, earlier theories have not discussed the role of deformations in determining the superconducting properties of BLG \cite{Uchoa,Hosseini1,Hosseini2,Hesselmann}. 

In this paper, we construct the Green's function for various coupling scenarios and present an extensive study of phonon-mediated pairing possibilities in BLG. As mentioned above, a recent experiment has demonstrated two phonon modes in a twisted BLG \cite{2ph-el_exp}. One mode was attributed to the intralayer vibrations while the other mode was related to interlayer couplings. Therefore, in the same spirit, one can expect that BLG can potentially develop two-gap superconductivity in the phonon-mediated electron-electron coupling picture, where each gap is mainly supported by one of the two phonon modes. Motivated by this picture, we also develop a two-gap superconductivity model for BLG. We employ the effective Hamiltonian method, consider a generic coupling matrix for the interlayer coupling potential of BLG, and study the different superconducting pairings and correlations. The generic expressions we derive can be applied to both twisted and shifted BLG when using the appropriate coupling terms, as will be discussed below.

For concreteness, we study shifted BLG with specific shift directions so that AA stacking turns into AB, AB into AC, and AC into AA in the extreme limits of the shift. We find that for an undoped system, the anomalous Green's function possesses only even-parity $s$-wave class and odd-parity $p$-wave class of symmetries for both AA and AB orderings with intralayer/interlayer phonon-mediated electron-electron coupling. For a doped system, however, AB ordering offers a rich variety of pairing symmetries such as $s$-wave, $p$-wave, $d$-wave, $f$-wave, and symmetry combinations. On the other hand, AA retains its previous even-parity $s$-wave, odd-parity $p$-wave (now $f$-wave class) symmetries. Introducing a generic coupling potential that accommodates in-plane twist and shift between layers, a $d$-wave symmetry can appear for both AA and AB orderings at $\mu=0$. Therefore, our findings suggest that by shifting the layers of a BLG with respect to each other, an effective switching to $d$-wave superconducting pairing with specific pseudospin (Pspin) states is accessible. This phenomenon offers a platform to control both superconducting critical temperature and the Pspin degree of freedom of superconducting correlations by experimentally simple actions that can find crucial applications in spintronics in addition to great interest to fundamental researches. The pairing symmetries predicted here can be probed experimentally by point contact tunneling spectroscopy experiments so that the conductance in different directions will be different and reveals these symmetries. It is worth mentioning that similar theory study as the present paper was conducted for black phosphorus monolayer in Ref. \onlinecite{BP2019}. It was found that a significant transition from effective $s$-wave ($p$-wave) to $d$-wave ($f$-wave) symmetry class is accessible simply by the exertion of strain into the plane of black phosphorus monolayer.

Using the obtained anomalous Green's function, we also calculate the intralayer and interlayer superconducting gaps as a function of temperature. For a given $s$-wave electron-electron interaction potential, our results show that the intralayer coupling in AA stacking order has both the largest gap amplitude and critical temperature, while the next largest gap and critical temperature belongs to interlayer coupling in AB stacking order. Also, the shift between the layers has opposite effects on the gap and critical temperature of intralayer and interlayer superconductivity in AA and AB orderings: the displacement applied to AA stacking with intralayer coupling or AB stacking with interlayer coupling reduces both the gap and critical temperature, while it enhances both of them for AA stacking with interlayer coupling and for AB stacking with intralayer coupling. 

We note that our analysis include the influence of chemical potential (away from the charge neutrality point) and are independent of the amplitude of the phonon-mediated electron-electron interaction: $\Delta$. We emphasize that our paper discusses the possible symmetries of superconducting pair correlations away from the magic angle of an incommensurately twisted BLG.

In Sec. \ref{sec1}, we first outline the Green's function technique, the effective Hamiltonian approach, and discuss different pairing scenarios that may arise in BLG (some related discussions are presented in Appendix \ref{apdx1}). In Sec. \ref{sec2}, we study the anomalous Green's function both analytically and numerically for different scenarios in AA and AB stacking orders. We first consider an undoped system and derive analytic expressions to the components of anomalous Green's function in Sec. \ref{subsec21}. Next, for a doped system, we evaluate the Green's functions numerically, and support these results by analytic expressions given in Appendix \ref{apdx2}. In Sec. \ref{sec3}, we consider a generic coupling potential between the two pristine graphene layers and repeat our studies of Sec. \ref{sec1}, now in the presence of a small displacement between the two layers. We also study the temperature dependencies of the superconducting gaps. We finally give concluding remarks in Sec. \ref{sec4}.

\section{Low energy effective Hamiltonian and green's function approach}\label{sec1}

We follow the effective Hamiltonian strategy \cite{ray1,ray2,ray_slip} where the Hamiltonian governing the low-energy physics of a generic BLG involving an arbitrary in-plane displacement between the layers can be expressed as
\begin{eqnarray}\label{Hamil1}
H&=& \int \frac{d\textbf{k}}{(2\pi)^2}\hat{\psi}^\dag(\textbf{k})H(\textbf{k})\hat{\psi}(\textbf{k})\\
&=&\int \frac{d\textbf{k}}{(2\pi)^2}\hat{\psi}^\dag(\textbf{k})\Big\{  H_1({\bf k})\rho_1+H_2({\bf k})\rho_2 + \tilde{\bm T}(\textbf{k})\rho_+ + \tilde{\bm T}^\dag(\textbf{k})\rho_-   \Big\}\hat{\psi}(\textbf{k}),\nonumber\label{heff}
\end{eqnarray}
in which the layer degree of freedom is described in terms of Pauli matrices $\rho_{0,z,x,y}$, and we have defined $2\rho_1=\rho_0+\rho_z$, $2\rho_2=\rho_0-\rho_z$, $2\rho_+=\rho_x+i\rho_y$, $2\rho_-=\rho_x-i\rho_y$.  The matrix $\tilde{\bm T}(\textbf{k})$ couples the top ($1$) and bottom ($2$) pristine single-layer graphene (SLG) and contains all information about the relative displacement between the layers (see below) \cite{ray1,ray2,ray_slip}. The Hamiltonians of the top and bottom layers are denoted by $H_{1,2}({\bf k})=\hbar v_\textbf{F}{\bf k}\cdot {\bm \sigma}$, respectively, in which ${\bm \sigma}=(\sigma_x,\sigma_y)$ and $\sigma_x, \sigma_y$ are Pauli matrices. The multiplication of Pauli matrices $\rho_i$ and $\sigma_i$ implies a tensor product so that the Hamiltonian (\ref{Hamil1}) is a $4\times 4$ matrix and the matrix $\tilde{\bm T}(\textbf{k})$ has an implicit $\bm \sigma$ (refer to Eqs. (\ref{tmatrix}) and (\ref{tmatrix2}) for further details). Here, the two-dimensional momentum \textbf{k} is in the plane of SLG. The field operator associated with the Hamiltonian is given by $\hat{\psi}^\dag(\textbf{k})=(\psi_{1\ua}^\dag,\psi_{1\da}^\dag,\psi_{2\ua}^\dag,\psi_{2\da}^\dag)$ where the first index is 1 (2) for the top (bottom) SLG and the second index is $\uparrow$ ($\downarrow$) for Pspin up -sublattice A- (Pspin down -sublattice B-). Note that, according to the mean-field definition of BCS superconductivity, two electrons with opposite spins and momenta are coupled through an attractive potential. This means that electron and hole excitations should be taken from opposite corners of the Brillouin zone \cite{RMP-2008-Beenakker, RMP-2009-Neto,PRL-2013-Halter}. In this case, one can show that the system possesses spin degeneracy \cite{RMP-2008-Beenakker, RMP-2009-Neto,PRL-2013-Halter}.  Nonetheless, to denote the Pspin and keep our notation simplified simultaneously, we have dropped the spin and valley indices. A detailed discussion is presented in Appendix \ref{apdx1}.

\begin{figure}[t]
\includegraphics[clip, trim=0.5cm 0.0cm 0.4cm 0.2cm, width=8.50cm,height=6.70cm]{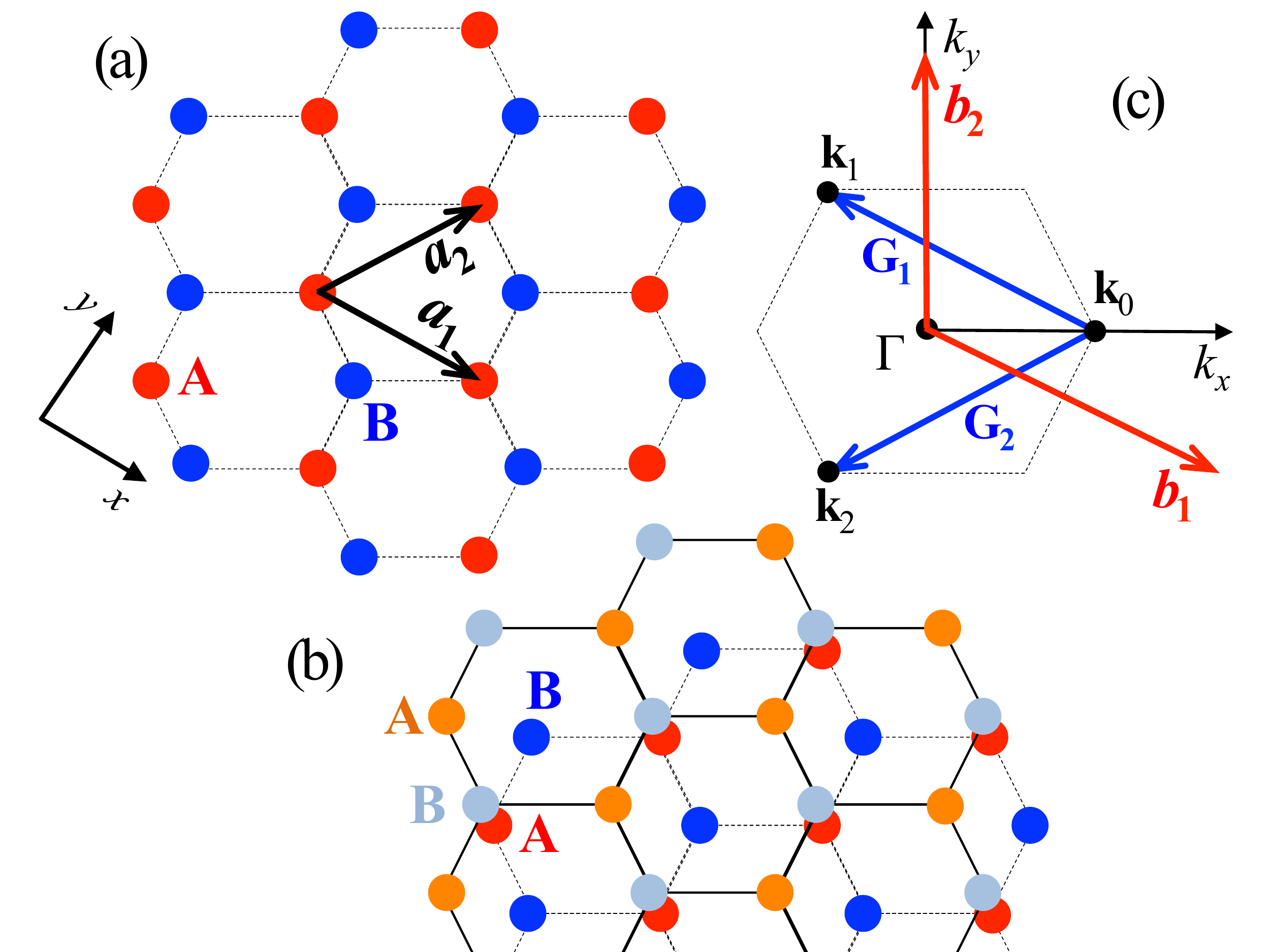}
\caption{\label{model} (Color online). (a) Graphene monolayer in real space (circles stand for carbon atoms) that is described by primitive vectors ${\bm a}_1$ and ${\bm a}_2$. The plane of graphene is located in the $xy$ plane and the honeycomb lattice can be separated into two sublattices A and B. (b) Illustration of an ideal AB stacking order where only the sublattice A of bottom layer (marked by red circles) is aligned with the sublattice B of top layer (marked by light blue circles). (c) Illustration of $\textbf{K}_j=\textbf{K}_0+\textbf{G}_j$ points in reciprocal space with primitive vectors ${\bm b}_1$ and ${\bm b}_2$ used in the calculations.}
\end{figure}

We next define the parameters that characterize a BLG system with a relative displacement between the top and bottom SLG. The monolayer pristine graphene can be divided into two sublattices A and B that can be described by two sublattice positions in real space ${\bm \nu}_A = (0,0), {\bm \nu}_B =2({\bm a}_1+{\bm a}_2)/3$ and two lattice vectors ${\bm a}_1 = a(1,0), {\bm a}_2 = a(1,\sqrt{3})/2$. Using this notation, the different stackings can be described by the sublattice positions given in Table \ref{table1}. Figure \ref{model}(a) displays graphene lattice in real space with sublattices A and B marked by dark red and dark blue circles, respectively. Also, we have illustrated AB ordering in Fig. \ref{model}(b) where only the sublattice B of top monolayer graphene is aligned with the sublattice A of bottom layer. In AA case, A and B sublattices of the top layer are aligned with A and B sublattices of the bottom layer (see Table \ref{table1}).
\begin {table}[t!]
\caption {Sublattice positions of BLG for different stacking orders. } \label{table1}
\begin{center}
\begin{tabular}{ |c|c|c|c| }
\hline
\hline
Stacking$\rightarrow$ & AA & AB & AC \\
 \hline
 \hline
\multirow{1}{*}
 {${\bm \nu}_{1,A}=$} & (0,0) & (0,0) & (0,0) \\
 \hline
 {${\bm \nu}_{1,B}=$} & {$2({\bm a}_1+{\bm a}_2)/3$} & {$2({\bm a}_1+{\bm a}_2)/3$} & {$2({\bm a}_1+{\bm a}_2)/3$} \\
  \hline
  \hline
\multirow{1}{*}
  {${\bm \nu}_{2,A}=$} & (0,0) & {$({\bm a}_1+{\bm a}_2)/3$} & {$2({\bm a}_1+{\bm a}_2)/3$}  \\
  \hline
 {${\bm \nu}_{2,B}=$} & {$2({\bm a}_1+{\bm a}_2)/3$} & (0,0) & {$({\bm a}_1+{\bm a}_2)/3$}\\
 \hline
 \hline
\end{tabular}
\end{center}
\end{table}
The reciprocal lattice of a BLG can be described by the lattice vectors ${\bm b}_1 = 2a^{-1}\pi(1,-1/\sqrt{3}), {\bm b}_2 = 2a^{-1}\pi(0,2/\sqrt{3})$ in reciprocal space as shown in Fig. \ref{model}(c). Since we are interested in the low-energy physics, the Hamiltonian can be expanded around reciprocal sublattice points:
\begin{eqnarray}
\textbf{K}_j=\textbf{K}_0+\textbf{G}_j,
\end{eqnarray}
in which $\textbf{K}_0=(2{\bm b}_1+{\bm b}_2)/3, \textbf{G}_0={\bm 0}, \textbf{G}_1=-{\bm b}_1, \textbf{G}_2=-{\bm b}_1-{\bm b}_2$ (see Fig. \ref{model}(b)).
The periodic arrangement of atoms in SLG is spanned by $\mathbf{ r}=\textbf{R}_j+{\bm \nu}_\alpha$, where $\textbf{R}_j$ are integer combinations of the primitive lattice vectors ${\bm a}_j$, and the orbitals centered at positions ${\bm \nu}_\alpha$. In the following, $\mathbf{ r}$ is the location at which the coupling between the top and bottom layers takes place. For example, the coupling of A and B sublattices shown in Fig. \ref{model}(b). The displacement enters the coupling matrix $\tilde{\bm T}$ via a phase factor \cite{ray1,ray2,ray_slip},
\begin{equation}\label{tmatrix}
\tilde{\bm T}(\textbf{k})=\sum_{j=0,1,2}\text{M}_j^\text{XX} \frac{t_{\perp}(\textbf{K}_j+\textbf{k})}{3}e^{i(\textbf{K}_j+\textbf{k})(\textbf{u}_{2}-\textbf{u}_{1})},
\end{equation}
in which $t_{\perp}(\textbf{K}_j+\textbf{k})$ is the interlayer hopping amplitude, given by $t_{\perp}({\bf q})=V_u^{-1}\int d{\bf r} t_{\perp}({\bf r})e^{i{\bf q}{\bf r}}$, $t_{\perp}({\bf r})=\sum_{i,j}t_{2,j}^{1,i}\Big\langle {\bf r},2\Big|c^\dag_{2,j}c_{1,i}\Big|{\bf r}',1\Big\rangle$ with $c^\dag,c$ being the creation and annihilation operators and ${\bf r}'={\bf r}+{\bm \delta}$ in which ${\bm \delta}$ denotes the distance between hopping sites. $V_u$ is the volume of unit cell, $i,j$ run over the lattice sites, $\textbf{u}_{1,2}$ are the displacements of the top and bottom layers, and $\text{M}_j^\text{XX}$ defines stacking order:
\begin{eqnarray}\label{tmatrix2}
&\text{M}_j^\text{AA}=\left( \begin{array}{cc}
1 & e^{-i(2j\pi/3)}\\
e^{+i(2j\pi/3)} & 1
\end{array}\right),\;\;\; \text{M}_j^\text{AB}=\left( \begin{array}{cc}
e^{+i(2j\pi/3)} & 1\\
e^{-i(2j\pi/3)} & e^{+i(2j\pi/3)}
\end{array}\right),\nonumber \\& \text{M}_j^\text{AC}=\left( \begin{array}{cc}
e^{-i(2j\pi/3)} & e^{+i(2j\pi/3)}\\
1 & e^{-i(2j\pi/3)}
\end{array}\right).
\end{eqnarray}

In what follows, we consider two types of couplings that may lead to phonon-mediated superconductivity in a BLG \cite{2ph-el_exp}. The top and bottom SLG are assumed to possess identical phonon modes while the coupling between the two SLG induces a new and different phonon mode \cite{2ph-el_exp}. Thus, the BCS electron coupling within each layer is called `intralayer phonon-mediated electron-electron coupling' and the electron coupling between the two layers is called `interlayer phonon-mediated electron-electron coupling' throughout this paper. In this basis, the two types of pairing mechanisms are characterized by the following two-electron amplitudes: 
\begin{subequations}
\begin{eqnarray}
& i)\quad  &\Delta_\text{S}\Big\langle\psi^\dag_{1\ua}\psi^\dag_{1\da}\Big\rangle+\text{h.c.}\; {\rm and} \;  \Delta_\text{S}\Big\langle\psi^\dag_{2\ua}\psi^\dag_{2\da}\Big\rangle+\text{H.c.},\; \label{g11}\\
& ii)\quad &\Delta_\text{B}\Big\langle\psi^\dag_{1\ua}\psi^\dag_{2\da}\Big\rangle+\text{H.c.}\; ,\label{g12}
\end{eqnarray}
\end{subequations}
where $\Delta_\text{S}$ and $\Delta_\text{B}$ are the BCS spin-singlet phonon-mediated electron-electron coupling within each layer and between the layers, respectively. For more details see Appendix \ref{apdx1}.

We next consider the Green's functions for the generic BLG system in the presence of superconductivity. The normal Green's functions $g$ and the anomalous Green's function $f$ are defined as follows; \cite{BP2019}
\begin{subequations}\label{GF_comps}
\begin{eqnarray}
&&{ g }_{\rho\rho'}^{\sigma\sigma'}(\tau,\tau'; \mathbf{ r}, \mathbf{ r}') = -\langle {\cal T}_{\tau}\psi_{\rho\sigma} (\tau,\mathbf{ r}') \psi_{\rho'\sigma'}^{\dag}(\tau',\mathbf{ r}')  \rangle,~~~~~~~\\
&&\underline{g}_{\rho\rho'}^{\sigma\sigma'}(\tau,\tau'; \mathbf{ r}, \mathbf{ r}') = - \langle {\cal T}_{\tau} \psi^{\dag}_{\rho\sigma} (\tau,\mathbf{ r}) \psi_{\rho'\sigma'}(\tau',\mathbf{ r}')  \rangle,~~~~~~~\\
&&{ f}_{\rho\rho'}^{\sigma\sigma'}(\tau,\tau'; \mathbf{ r}, \mathbf{ r}') = + \langle {\cal T}_{\tau} \psi_{\rho\sigma} (\tau,\mathbf{ r}) \psi_{\rho'\sigma'}(\tau',\mathbf{ r}')  \rangle,~~~~~~~\\
&&f_{\rho\rho'}^{\sigma\sigma'\dag}(\tau,\tau'; \mathbf{ r}, \mathbf{ r}') = + \langle {\cal T}_{\tau} \psi^{\dag}_{\rho\sigma} (\tau,\mathbf{ r}) \psi^{\dag}_{\rho'\sigma'}(\tau',\mathbf{ r}')  \rangle,~~~~~~~
\end{eqnarray}
\end{subequations}
where ${\cal T}_{\tau}$ is the time ordering operator, and $\tau, \tau'$ are the imaginary times. Here, $\sigma,\sigma'$ and $\rho,\rho'$ denote the Pspin and layer indices and $\langle ... \rangle$ is the thermodynamics averaging.
In particle-hole space, the Green's function satisfies:
\begin{eqnarray} \label{BCSH}
\left( \begin{array}{cc}
\hat{H}(\mathbf{ r})-i\omega_n& \hat{\Delta}(\textbf{r}) \\
\hat{\Delta}^*(\textbf{r}) &  {\cal T}_\text{t}\hat{H}(\mathbf{ r}){\cal T}_\text{t}^\dag+i\omega_n
\end{array} \right)\check{g}(i\omega_n;\mathbf{ r},\mathbf{ r}')
=\delta(\mathbf{ r}-\mathbf{ r}'),
\end{eqnarray}
in which $\omega_n=\pi (2n+1) k_B T $ is the Matsubara frequency, $n\in { Z}$, $k_B$ is the Boltzman constant, $ T $ is temperature, ${\cal T}_\text{t}$ is the time-reversal operator, and $\hat{\Delta}(\mathbf{ r})$ is the superconducting gap. Note that $\hat{H}(\mathbf{ r})$ is obtained by replacing ${\bf k}\rightarrow -i\boldsymbol{ \nabla}$ in $H(\textbf{k})$, in Eq. (\ref{Hamil1}). 
The matrix form of the Green's function is given by;
\begin{equation}
\check{g}(i\omega_n;\mathbf{ r},\mathbf{ r}')=\left(  \begin{array}{cc}
\hat{g}(i\omega_n;\mathbf{ r},\mathbf{ r}') & \hat{f}(i\omega_n;\mathbf{ r},\mathbf{ r}')\\
\hat{f}^\dag(i\omega_n;\mathbf{ r},\mathbf{ r}') & \hat{\underline{g}}(i\omega_n;\mathbf{ r},\mathbf{ r}')
\end{array}  \right).
\end{equation}
We denote $4\times 4$ matrices by `hat' symbol, $\hat{\square}$, and $8\times 8$ matrices by `check' symbol, $\check{\square}$.

We now proceed to derive the anomalous Green's function $f$ for the different coupling scenarios (Eq. (\ref{gapfunc}) below gives the self-consistency condition these functions must satisfy).\cite{BP2019}

\section{Bilayer graphene without any deformations}\label{sec2}

To begin, we consider spin-singlet intralayer and interlayer phonon-mediated electron-electron couplings and find possible superconducting pair correlations in a BLG with zero doping. We thus assume that the top and bottom SLG host identical phonon-mediated electron-electron interactions, i.e., $\Delta_\text{S}^1\equiv\Delta_\text{S}^2$. In what follows, we use a compact notation for the different combinations of momenta (see Table \ref{table2}) that arise during the calculations to simplify the resulting expressions. We first consider the simplest cases that result in analytic solutions and gradually add parameters to reach more complicated low-energy effective Hamiltonians.

\begin {table*}[t]
\caption {Different combinations of momentum functions, appearing in the calculations. } \label{table2}
\normalsize
\begin{center}
\begin{tabular}{|cccc|}
\hline
\hline
\multicolumn{1}{|c|} {$ k_+=k_x+ik_y$}   &  \multicolumn{1}{c|} {$k_+^2=k_x^2 + 2 i k_x k_y - k_y^2$} &  \multicolumn{1}{c|} {$k_+^3=k_x^3 + 3 i k_x^2 k_y - 3 k_x k_y^2 - i k_y^3$}    &  $k_+^4=k_x^4 + 4 i k_x^3 k_y - 6 k_x^2 k_y^2 - 4 i k_x k_y^3 + k_y^4$    \\
 \multicolumn{1}{|c|} {$k_-=k_x-ik_y$}  &  \multicolumn{1}{c|} {$k_-^2=k_x^2 - 2 i k_x k_y - k_y^2$}    &  \multicolumn{1}{c|} {$k_-^3=k_x^3 - 3 i k_x^2 k_y - 3 k_x k_y^2 + i k_y^3$}    &  $k_-^4=k_x^4 - 4 i k_x^3 k_y - 6 k_x^2 k_y^2 + 4 i k_x k_y^3 + k_y^4$     \\

 \multicolumn{1}{|c|} { $k_+k_-=k_x^2 + k_y^2$}  &   \multicolumn{1}{c|} {$k_+^2k_-^2=(k_x^2 +  k_y^2)^2$}   &  \multicolumn{1}{c|} {$k_+^3k_-^3=(k_x^2 +  k_y^2)^3$}    &   $k_+^4k_-^4=(k_x^2 +  k_y^2)^4$    \\
\hline
\multicolumn{2}{|c|} {$k_-^2k_+=k_x^3 - i k_x^2 k_y + k_x k_y^2 - i k_y^3$}

    &    \multicolumn{2}{c|} {$k_-^2k_+^3=k_x^5 + i k_x^4 k_y + 2 k_x^3 k_y^2 + 2 i k_x^2 k_y^3 + k_x k_y^4 + i k_y^5$}

\\
\multicolumn{2}{|c|} {$k_-^3k_+=k_x^4 - 2 i k_x^3 k_y - 2 i k_x k_y^3 - k_y^4$}   &   \multicolumn{2}{c|} {$k_++k_-=2 k_x$}    \\

\multicolumn{2}{|c|} {$k_-^3k_+^2=k_x^5 - i k_x^4 k_y + 2 k_x^3 k_y^2 - 2 i k_x^2 k_y^3 + k_x k_y^4 - i k_y^5$}   &  \multicolumn{2}{c|} {$k_+-k_-=2 i k_y$}    \\
\hline
\hline
\end{tabular}
\end{center}
\end{table*}

This section is split into two subsections, where we study the undoped and doped cases, respectively, considering both AA and AB stacking orders. For symmetry reasons, the AC stacking order leads to similar results as AB stacking and, thus, we omit presenting results for the AC stacking order. We also assume that the system is translationally invariant in the $xy$ plane (the plane of BLG) and thus the corresponding Hamiltonian is diagonal in momentum space as a function of $k_x$ and $k_y$. Consequently, we deal with a set of algebraic equations in momentum space for the Green's function. The anomalous components of Green's function ${ f}_{\rho\rho'}^{\sigma\sigma'}$ correspond to pairing correlations that may arise in the system whereas the normal Green's function ${ g}_{\rho\rho'}^{\sigma\sigma'}$ contains information about local density of states. Hence, in what follows, we focus on the anomalous Green's function ${ f}_{\rho\rho'}^{\sigma\sigma'}$ and do not present explicitly the normal Green's function ${ g}_{\rho\rho'}^{\sigma\sigma'}$ although we have derived all the components in our actual calculations. 

In the following, inspired by the atomic orbital symbols, we use the following nomenclature: terms proportional to $k_\pm$(=$k_x\pm ik_y$) $\rightarrow$ $p$-wave, $k_\pm^2$ $\rightarrow$ $d$-wave, and $k_\pm^2k_\mp$ $\rightarrow$ $f$-wave, etc. (see also Tables \ref{table2} and \ref{table3}).

In Secs. \ref{subsec21} and \ref{subsec22}, we consider undoped $\mu=0$ and doped $\mu\neq 0$ BLG, respectively. Each of these sections studies both AA and AB orderings with both intralayer $\Delta_\text{S}$ and interlayer $\Delta_\text{B}$ phonon-mediated electron-electron interaction. 

\subsection{Undoped bilayer graphene: $\mu=0$}\label{subsec21}

We consider two cases separately: spin singlet intralayer phonon-mediated electron-electron coupling $\Delta_\text{S}$ and spin-singlet interlayer phonon-mediated electron-electron coupling $\Delta_\text{B}$. In both cases, we assume a constant coupling potential between the two monolayers given by: $t=t^\dag$. The components of coupling potential are defined by Eq. (\ref{tmatrix}). 

\subsubsection{Intralayer coupling: $\Delta_\text{S}\ne 0, \Delta_\text{B}=0$}\label{mu0intra}

$\bullet$\textbf{AA stacking}:
In this case, the components of the anomalous Green's function Eqs. (\ref{GF_comps}) are: 
\begin{subequations}
\begin{eqnarray}
&&\Omega f_{11}^{\ua\ua}= 0,\\
&&\Omega f_{11}^{\ua\da}=-\Delta_\text{S}(k_-k_++t^2+\Delta_\text{S}^2+\omega_n^2), \\
&&\Omega f_{12}^{\ua\ua}=-2t\Delta_\text{S} k_-, \\
&&\Omega f_{12}^{\ua\da}=0, \\
&&\Omega f_{11}^{\da\ua}=+\Delta_\text{S}(k_-k_++t^2+\Delta_\text{S}^2+\omega_n^2), \\
&&\Omega f_{11}^{\da\da}=0, \\
&&\Omega f_{12}^{\da\ua}=0, \\
&&\Omega f_{12}^{\da\da}=+2t\Delta_\text{S} k_+\\
&&\Omega=k_-^2k_+^2+2k_-k_+(\Delta_\text{S}^2-t^2+\omega_n^2)+(t^2+\Delta_\text{S}^2+\omega_n^2)^2. 
\end{eqnarray}
\end{subequations}
As seen, the intralayer Pspin-triplet and interlayer Pspin-singlet correlations are not induced while, as expected, the intralayer Pspin-singlet pairings are nonzero and have even-parity even-frequency symmetry. Interestingly, the Pspin-triplet interlayer correlations are nonzero, proportional to the coupling strength between the layer and possess odd-parity even-frequency symmetry.
\\*
$\bullet$\textbf{AB stacking}: Next, we consider AB stacking order and intralayer opposite-spin phonon-mediated electron-electron interactions. Constructing the anomalous Green's function, we find: 
\begin{subequations}
\begin{eqnarray}
&&\Omega f_{11}^{\ua\ua}= 0, \\
&&\Omega f_{11}^{\ua\da}=-\Delta_\text{S}(k_-k_++\Delta_\text{S}^2+\omega_n^2), \\
&&\Omega f_{12}^{\ua\ua}=- it\Delta_\text{S}\omega_n, \\
&&\Omega f_{12}^{\ua\da}=0, \\
&&\Omega f_{11}^{\da\ua}=+\Delta_\text{S}(k_-k_++\Delta_\text{S}^2+\omega_n^2), \\
&&\Omega f_{11}^{\da\da}=0, \\
&&\Omega f_{12}^{\da\ua}=t\Delta_\text{S} (k_- - k_+), \\
&&\Omega f_{12}^{\da\da}=+it\Delta_\text{S}\omega_n, \\
&&\Omega=(k_-k_++\Delta_\text{S}^2)^2+(2k_-k_++t^2+2\Delta_\text{S}^2)\omega_n^2+\omega_n^4. 
\end{eqnarray}
\end{subequations}
Similar to the AA case, the intralayer Pspin-triplet correlations are zero while the Pspin-singlet pairings are nonzero with even-parity even-frequency. Due to the specific coupling matrix in the AB stacking order, one of the Pspin-singlet correlation components is odd-parity and even in frequency while the Pspin-triplet ones are even in parity and odd in frequency. Note that these unconventional correlations are proportional to the interlayer coupling strength $t$.

\subsubsection{Interlayer BCS coupling: $\Delta_\text{S}=0, \Delta_\text{B}\ne 0$}

Here, we consider a situation where one electron from the top layer and one from the bottom layer are coupled through the interlayer phonons which is given by Eq. (\ref{g12}). For AA and AB stack orderings of the two monolayer graphene, we find the following anomalous components of the Green's function, Eqs. (\ref{GF_comps});
\\*
$\bullet$\textbf{AA stacking}: 
\begin{subequations}
\begin{eqnarray}
&&\Omega f_{11}^{\ua\ua}= 0, \\
&&\Omega f_{11}^{\ua\da}=-2it\Delta_\text{B} \omega_n, \\
&&\Omega f_{12}^{\ua\ua}=0, \\
&&\Omega f_{12}^{\ua\da}=-\Delta_\text{B} (k_-k_+-t^2+\Delta_\text{B}^2+\omega_n^2), \\
&&\Omega f_{11}^{\da\ua}=2it\Delta_\text{B} \omega_n, \\
&&\Omega f_{11}^{\da\da}=0, \\
&&\Omega f_{12}^{\da\ua}=\Delta_\text{B} (k_-k_+-t^2+\Delta_\text{B}^2+\omega_n^2), \\
&&\Omega f_{12}^{\da\da}=0, \\
&&\Omega=(k_-k_+-t^2+\Delta_\text{B}^2)^2+2(k_-k_++t^2+\Delta_\text{B}^2)\omega_n^2+\omega_n^4.
\end{eqnarray}
\end{subequations}
\\*
$\bullet$\textbf{AB stacking}: 
\begin{subequations}
\begin{eqnarray}
&&\Omega f_{11}^{\ua\ua}= 0, \\
&&\Omega f_{11}^{\ua\da}=-t\Delta_\text{B} k_+, \\
&&\Omega f_{12}^{\ua\ua}=0, \\
&&\Omega f_{12}^{\ua\da}=-\Delta_\text{B} (k_-k_++\Delta_\text{B}^2+\omega_n^2), \\
&&\Omega f_{11}^{\da\ua}=t\Delta_\text{B} k_+, \\
&&\Omega f_{11}^{\da\da}=0, \\
&&\Omega f_{12}^{\da\ua}=\Delta_\text{B} (k_-k_++t^2+\Delta_\text{B}^2+\omega_n^2), \\
&&\Omega f_{12}^{\da\da}=0, \\
&&\Omega=k_-^2k_+^2+2k_-k_+(\Delta_\text{B}^2+\omega_n^2)+(\Delta_\text{B}^2+\omega_n^2)(\Delta_\text{B}^2+t^2+\omega_n^2).
\end{eqnarray}
\end{subequations}

Comparing with the Green's function components of the same stacking orders but with intralayer phonon-mediated electron-electron coupling discussed in Subsec. \ref{mu0intra}, we readily find that the AA stacking in the current scenario develops even-parity and odd-frequency superconducting correlations similar to the AB stacking order of Sec. \ref{mu0intra}. The same symmetry correspondence can be seen between AB stacking of this current subsection and Sec. \ref{mu0intra}. In that case, BLG develops odd-parity even-frequency correlations. Note that, unlike the intralayer phonon-mediated electron-electron coupling scenario, Sec. \ref{mu0intra}, here all components of the anomalous Green's function are spin-singlet. Also, the odd-parity components are of $p$-wave type (i.e., proportional to $k_{\pm}$. See Tables \ref{table2} and \ref{table3}). Thus, for the undoped case, the only possible pairing correlations with the intralayer and interlayer phonon-mediated electron-electron couplings are of $s$-wave and $p$-wave class regardless of the stacking order.

\begin{figure*}[thp]
\begin{tabular}{ll}
\includegraphics[width=9.0cm,height=7.80cm]{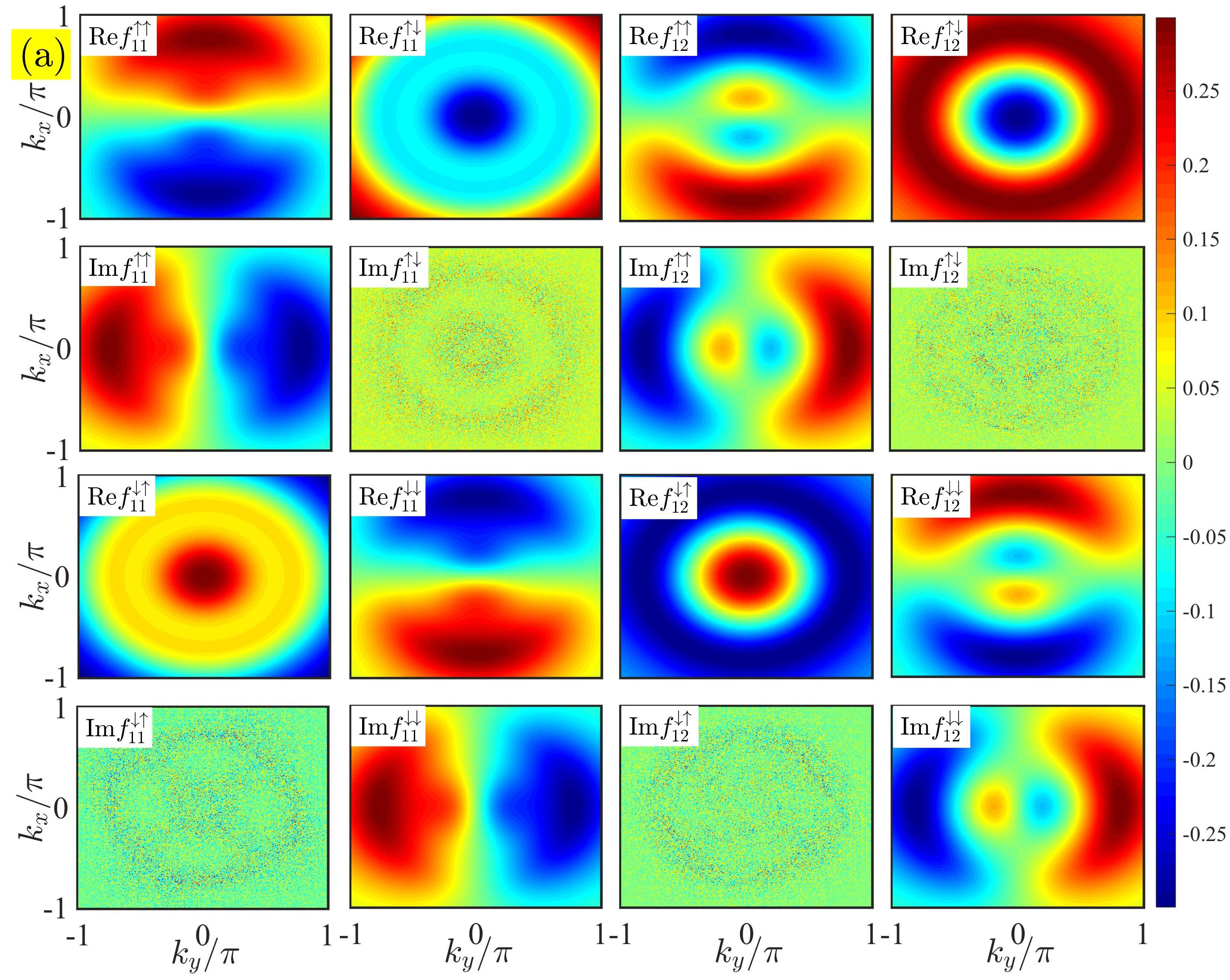}\hfill
&\hfill
\includegraphics[width=9.0cm,height=7.80cm]{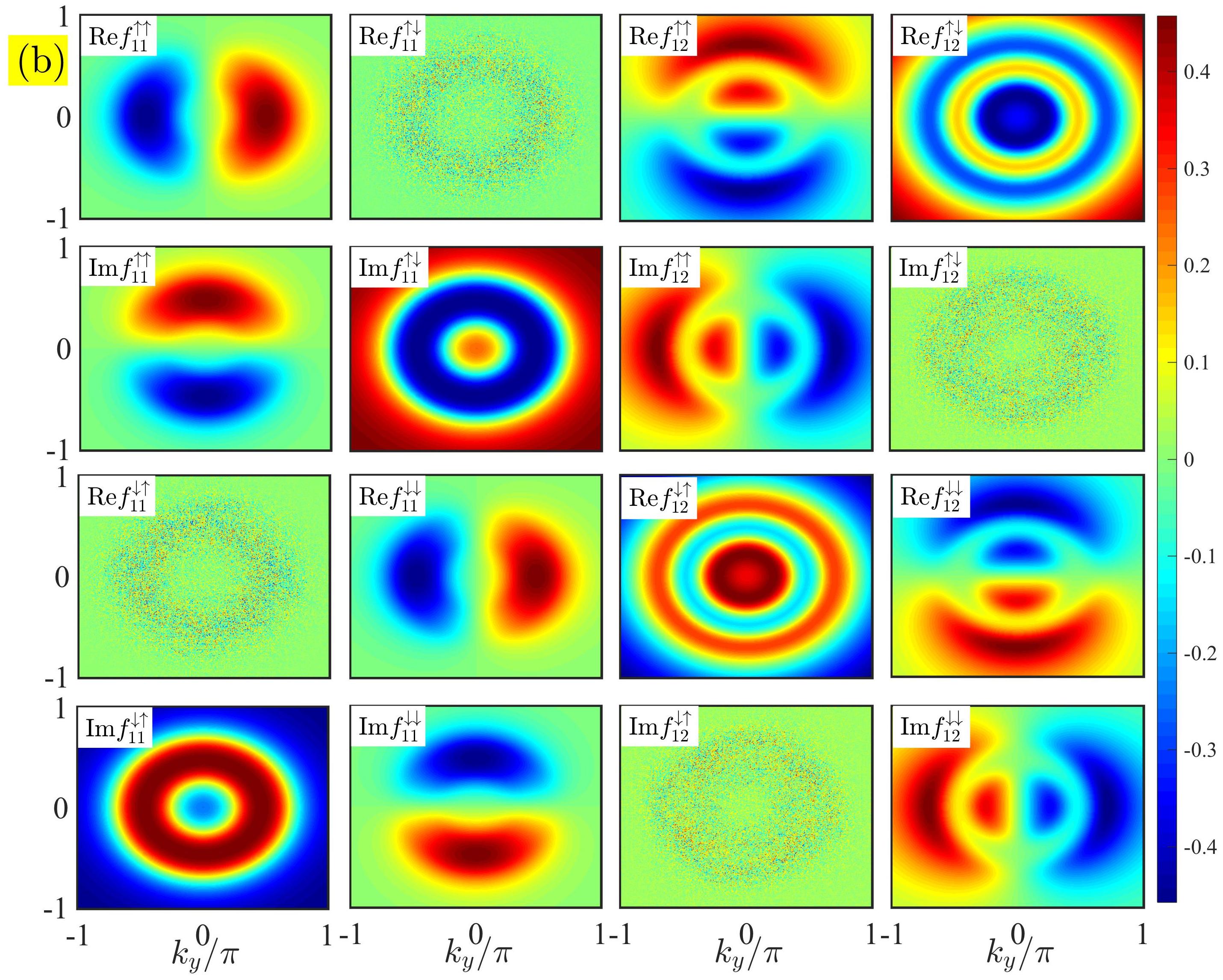}
\end{tabular}
\caption{ (Color online).
Real and imaginary parts of the different pairing amplitudes in a BLG with perfect AA stacking and finite doping $\mu\neq 0$. (a) Pure intralayer coupling: $\Delta_\text{S}\neq 0$ and $\Delta_\text{B}=0$. (b) Pure interlayer coupling: $\Delta_\text{S}= 0$ and $\Delta_\text{B}\neq 0$.  }\label{fig1}
\end{figure*}
\begin{figure*}[thp]
\begin{tabular}{ll}
\includegraphics[width=9.0cm,height=7.80cm]{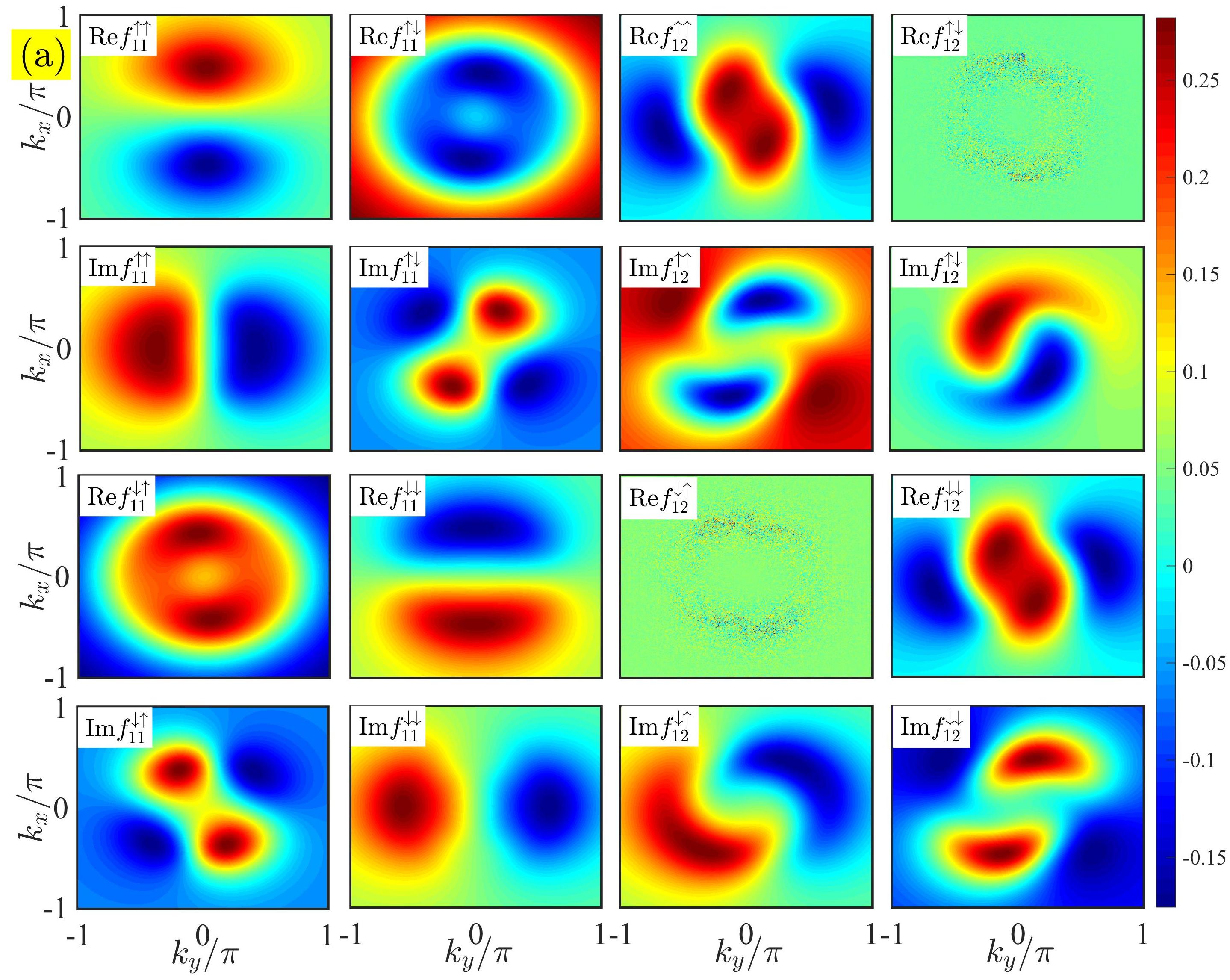}\hfill
&\hfill
\includegraphics[width=9.0cm,height=7.80cm]{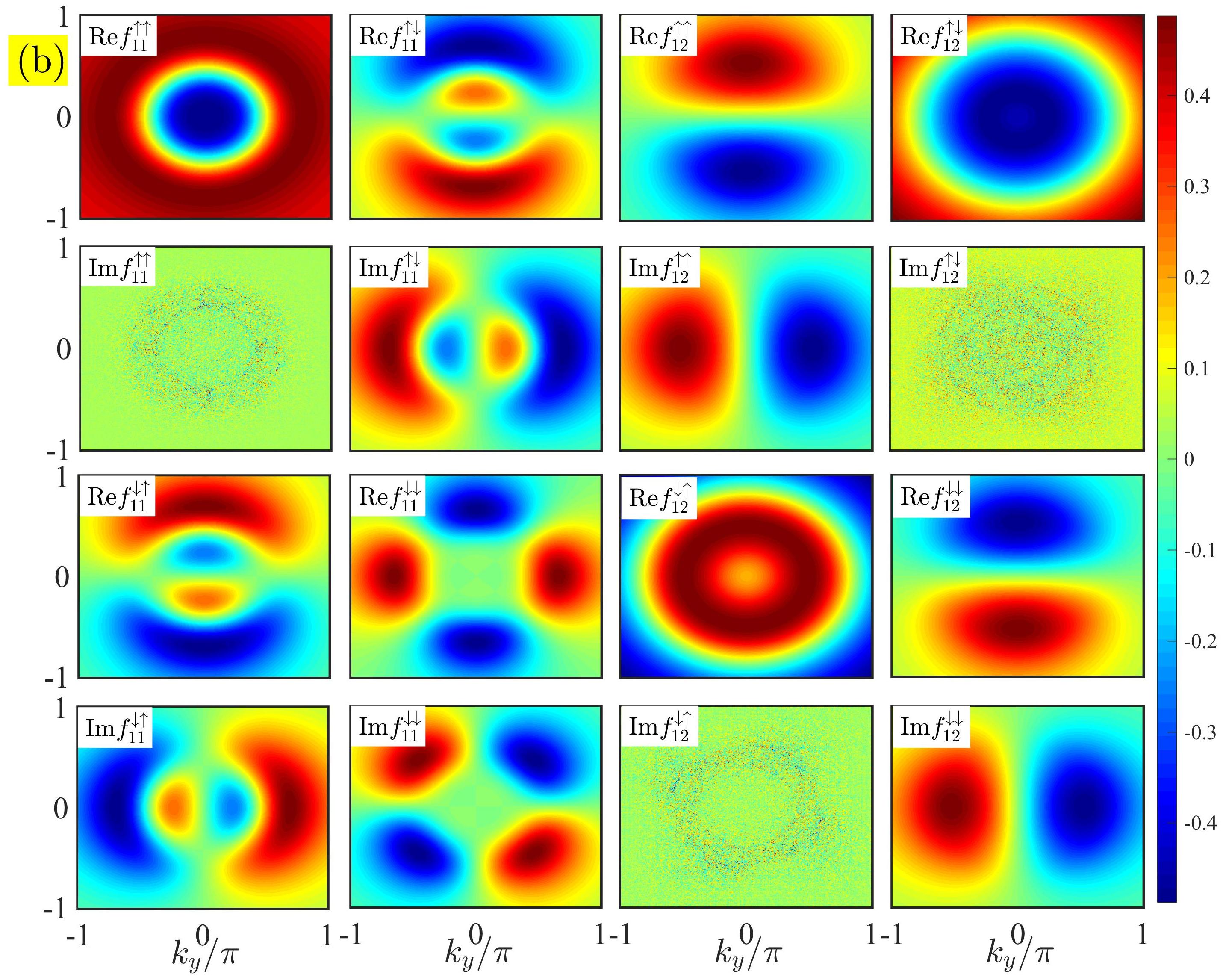}
\end{tabular}
\caption{ (Color online).
Real and imaginary parts of different pairings amplitudes in a BLG with a perfect AB stacking at finite doping $\mu\neq 0$. (a) Pure intralayer coupling: $\Delta_\text{S}\neq 0$ and $\Delta_\text{B}=0$. (b) Pure interlayer coupling: $\Delta_\text{S}= 0$ and $\Delta_\text{B}\neq 0$.  }\label{fig2}
\end{figure*}

\begin{figure*}[t]
\begin{tabular}{ll}
\includegraphics[width=9.0cm,height=7.80cm]{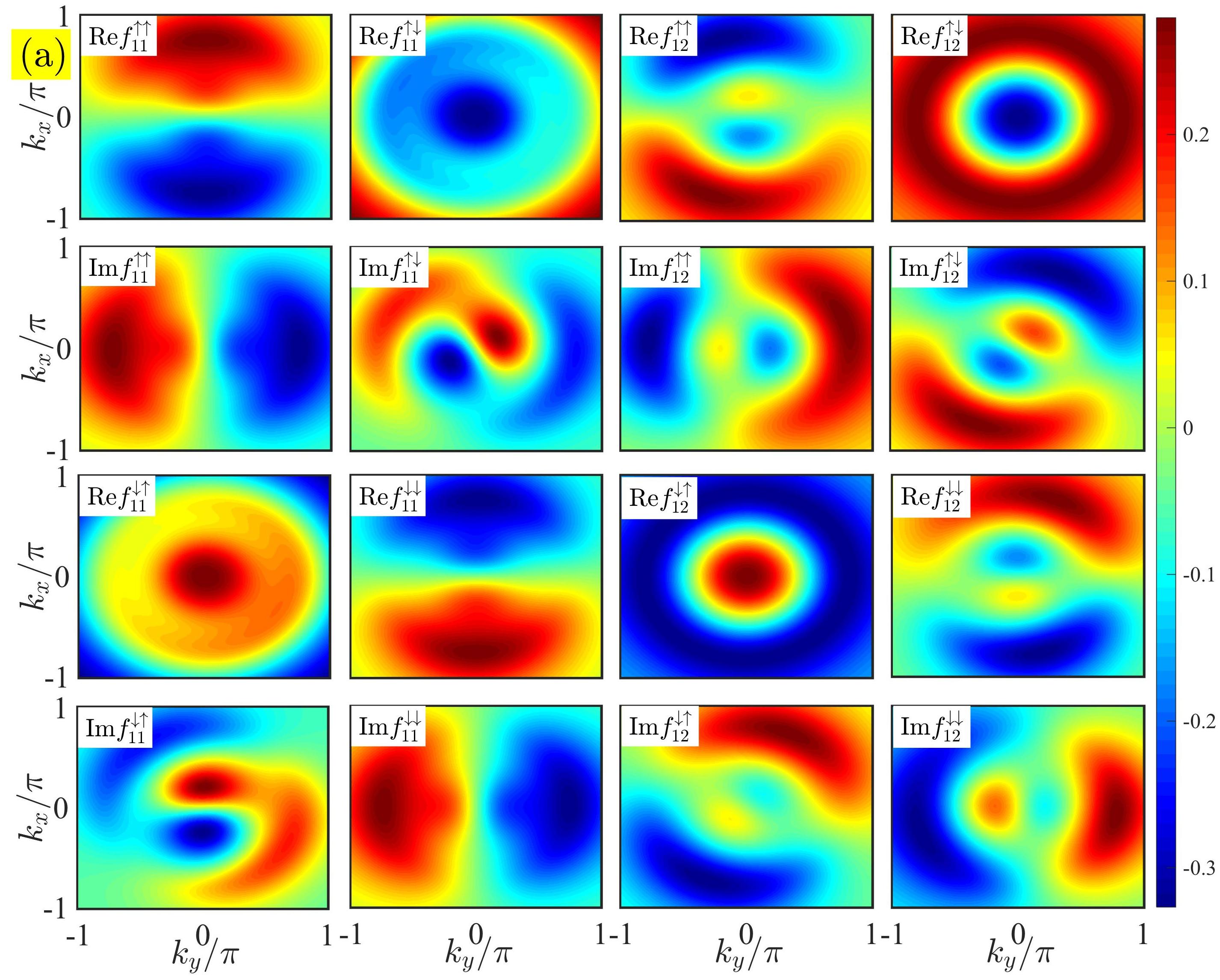}\hfill
&\hfill
\includegraphics[width=9.0cm,height=7.80cm]{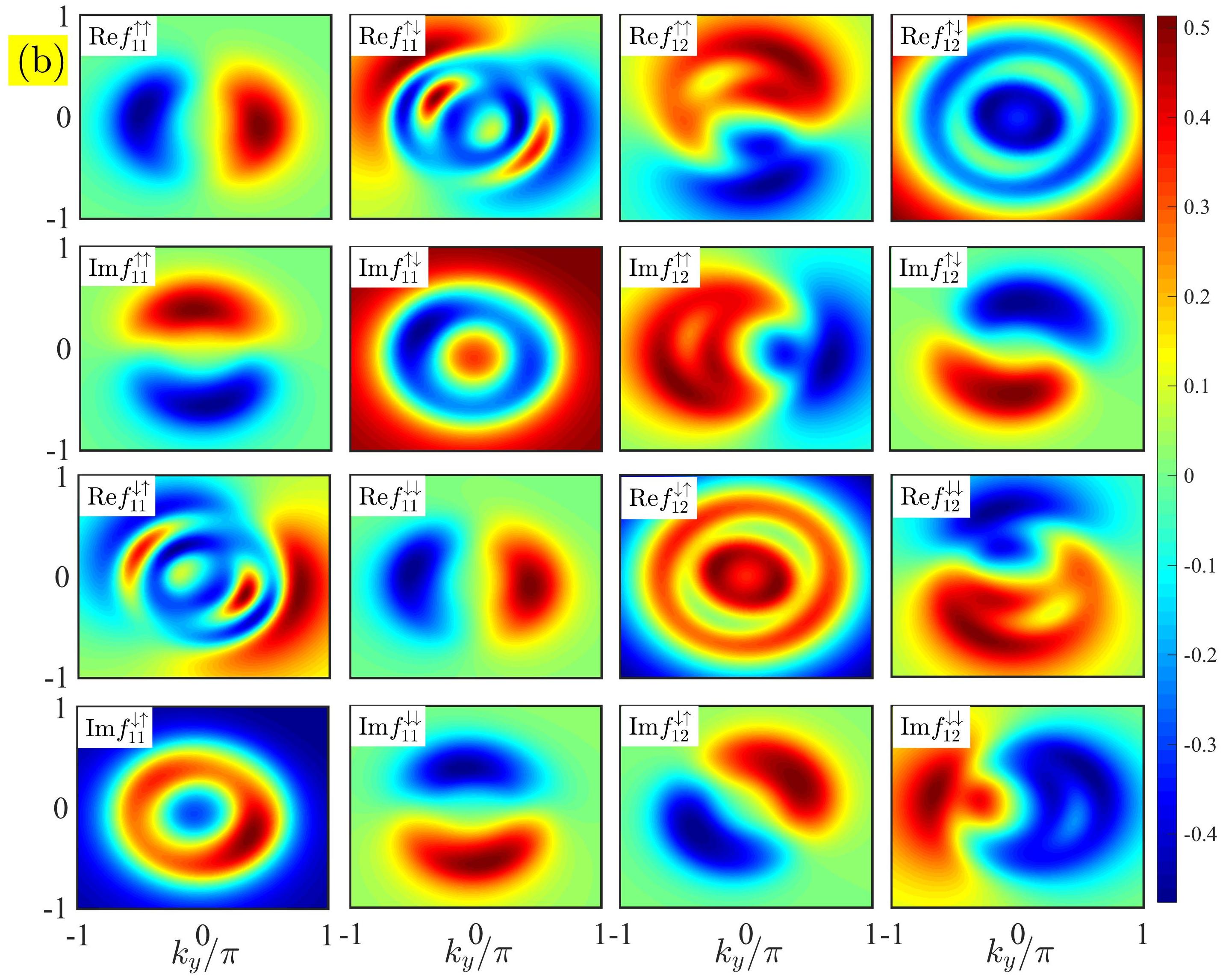}
\end{tabular}
\caption{ (Color online).
Real and imaginary parts of different pairings for a nonideal AA stacking configuration, i.e., in which the top SLG is shifted toward the AB configuration, $\epsilon=0.2$, at finite doping, $\mu\neq 0$ similar to the ideal AA stacking counterpart. (a) Purely intralayer coupling: $\Delta_\text{S}\neq 0$ and $\Delta_\text{B}=0$. (b) Pure interlayer coupling: $\Delta_\text{S}= 0$ and $\Delta_\text{B}\neq 0$.  }\label{fig3}
\end{figure*}
\begin{figure*}[t]
\begin{tabular}{ll}
\includegraphics[width=9.0cm,height=7.80cm]{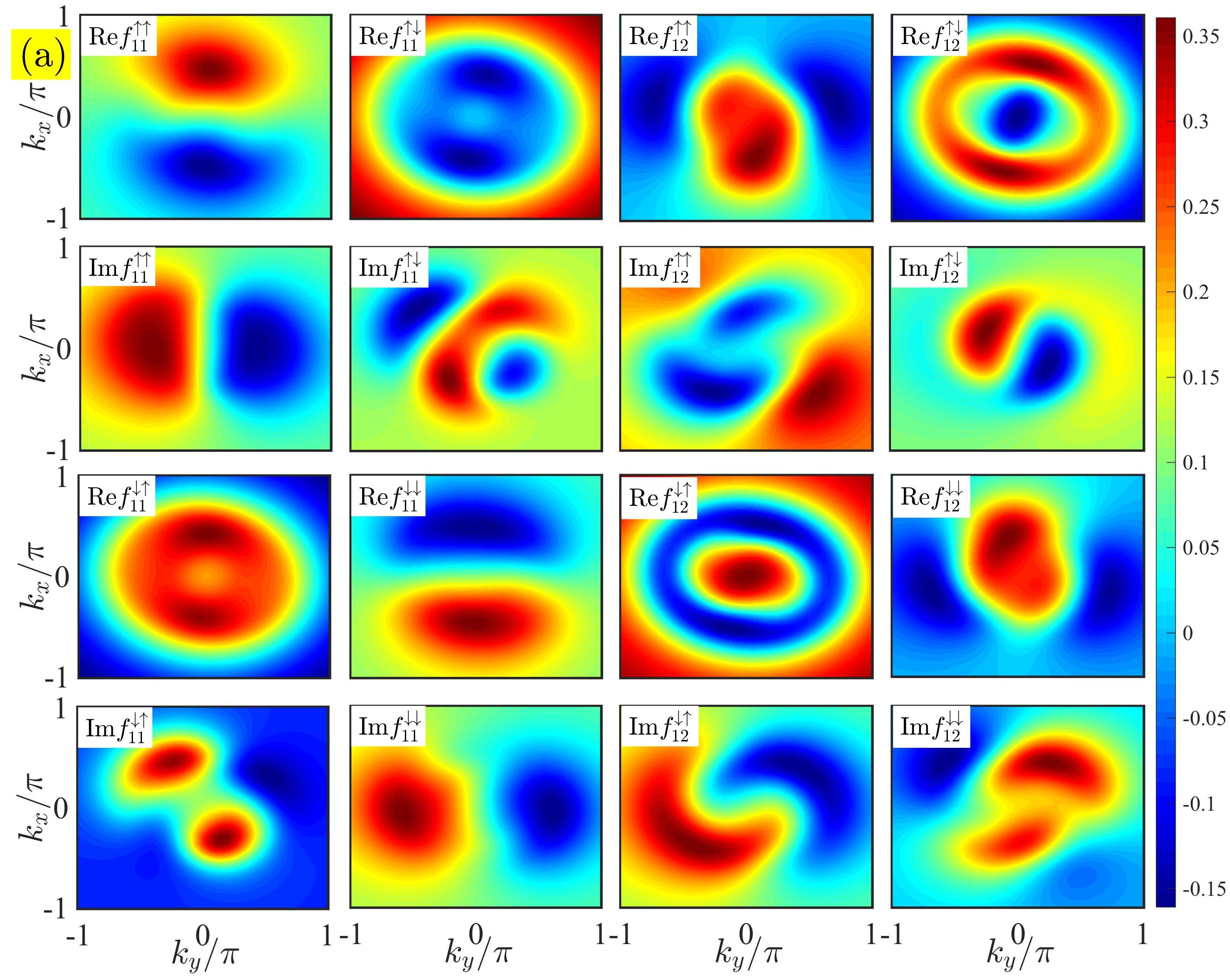}\hfill
&\hfill
\includegraphics[width=9.0cm,height=7.80cm]{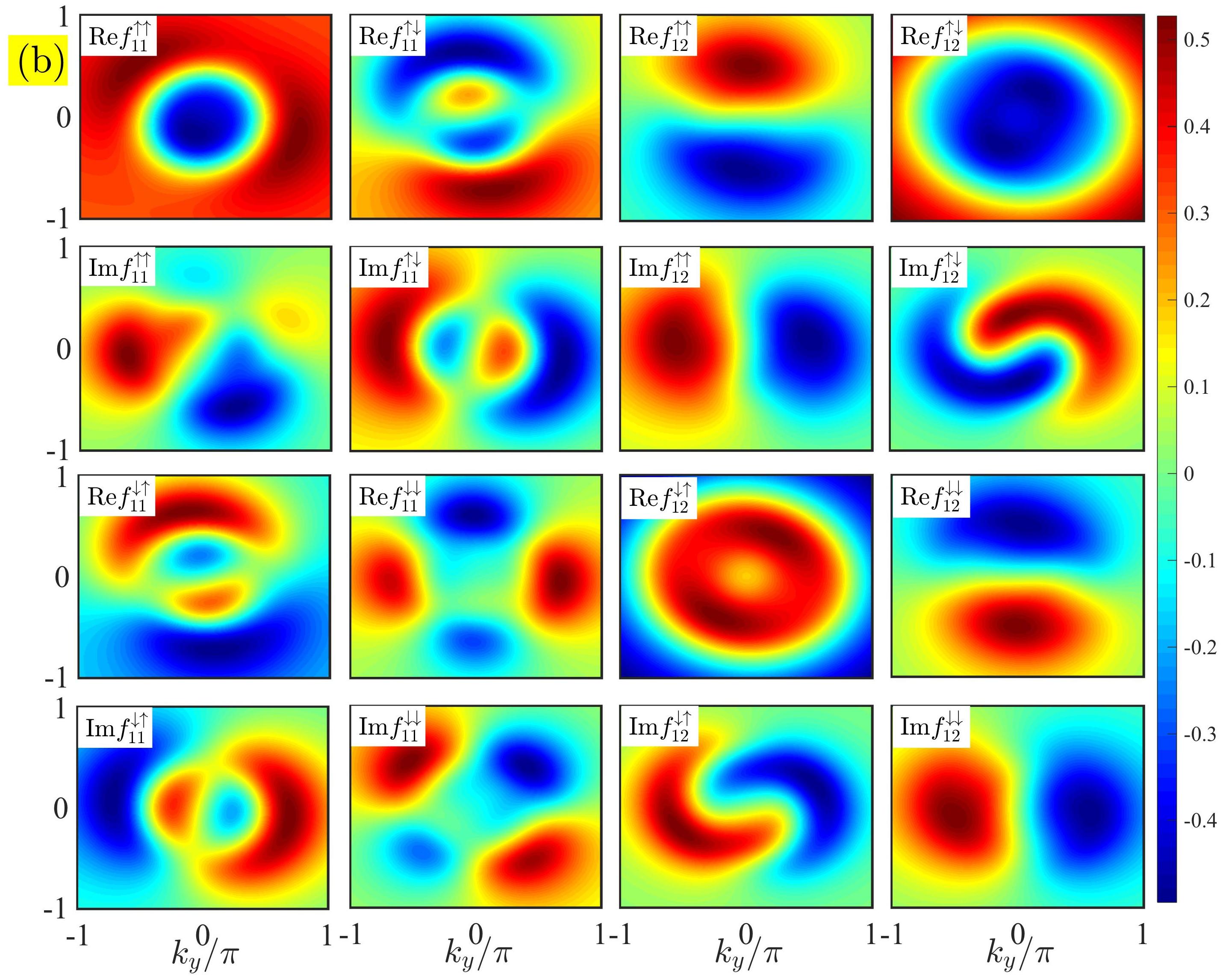}
\end{tabular}
\caption{(Color online).
Real and imaginary parts of different pairings for a nonideal AB stacking configuration, i.e., in which the top SLG is shifted toward AC stacking, $\epsilon=0.2$, at finite doping, $\mu\neq 0$ similar to the ideal AB stacking counterpart. (a) Pure intralayer coupling: $\Delta_\text{S}\neq 0$ and $\Delta_\text{B}=0$. (b) Pure interlayer coupling: $\Delta_\text{S}= 0$ and $\Delta_\text{B}\neq 0$.
 }\label{fig4}
\end{figure*}

\subsection{Bilayer graphene with finite doping $\mu\neq 0$}\label{subsec22}

Next, we consider finite doping, and reinvestigate the superconducting correlations given by the anomalous Green's function as discussed in the previous section. 

\subsubsection{\text{AA} stacking  order: phonon-mediated intralayer and interlayer electron-electron couplings}

In the presence of a finite doping, analytic expressions for the anomalous Green's function become cumbersome. However, they allow one to deduce and analyze the pairing symmetries that may arise. We have presented them in Appendix \ref{apdx2} and exploit them as a benchmark for our numerical calculations.

To evaluate the effective momentum dependencies, we set the Matsubara frequency fixed at the lowest mode with $n=0$ for the individual components of the anomalous Green's function evaluated at low temperatures (typically, $T\sim 0.01T_c$). We also set a `representative' ratio to the chemical potential over coupling strength $\mu/t=1.5$ throughout the numerical study without loss of generality. Figure \ref{fig1} illustrates the momentum space map of the real and imaginary parts of the Green's function components in AA stacking order. In Fig. \ref{fig1}(a), the phonon-mediated electron-electron coupling is of the intralayer type $\Delta_\text{S}\neq 0, \Delta_\text{B}=0$ while in Fig. \ref{fig1}(b) we consider only interlayer coupling $\Delta_\text{S}= 0, \Delta_\text{B}\neq 0$. For the intralayer coupling, Pspin-singlet $f_{11}^{\ua\da},f_{12}^{\ua\da},f_{11}^{\da\ua},f_{12}^{\da\ua}$ components are real and have $s$-wave ($\propto k_x^2+k_y^2$) symmetry while the other components with triplet Pspin state show $p$-wave ($\propto k_\pm$) and $f$-wave ($\propto k_\pm^2k_\mp$) symmetries with both nonzero imaginary and real parts. The interlayer coupling changes this picture slightly. Yet, the correlations with triplet Pspin state are of effective $f$-wave class while the intralayer Pspin-singlet correlations are purely imaginary with $s$-wave ($\propto k_x^2+k_y^2$) symmetry class. The interlayer Pspin-singlet correlations, $f_{12}^{\ua\da}$ and $f_{12}^{\da\ua}$, are of $s$-wave class and even in frequency. 

In short, the AA stacking order with and without doping and in the presence of intralayer or interlayer phonon-mediated electron-electron BCS coupling does not support $d$-wave symmetry class. Hence, AA stacking order can only yields $s$-wave, $p$-wave, and $f$-wave type superconducting correlations with the phonon-mediated scenarios considered in this paper. This can be fully confirmed by examining the analytic expressions given by Eqs. (\ref{intraAA}) and (\ref{interAA}) in Appendix \ref{apdx2}. According to the momentum combinations presented in Table \ref{table2}, Eqs. (\ref{intraAA}) and (\ref{interAA}) contain no term of type single $k_{\pm}^2$ that results in $d$-wave symmetry class. The allowed momentum combinations are either odd-parity or possess $s$-wave symmetry.

\subsubsection{\text{AB} stacking order: phonon-mediated intralayer and interlayer electron-electron couplings}

Figure \ref{fig2} shows the pairing correlations in a BLG with AB ordering of the top and bottom pristine layers. In Fig. \ref{fig2}(a), the BCS electron-electron coupling is of the intralayer type, i.e., $\Delta_\text{S}\neq 0, \Delta_\text{B}=0$, while Fig. \ref{fig2}(b) contains results with interlayer BCS electron-electron coupling, i.e., $\Delta_\text{S}= 0, \Delta_\text{B}\neq 0$. Clearly, in Fig. \ref{fig2}(a), components $f_{11}^{\ua\da},f_{12}^{\ua\ua},f_{11}^{\da\ua},f_{12}^{\da\da}$ show a combination of $d$-wave and $s$-wave symmetries both in imaginary and real parts. The rest of the pairing components are, however, of the $p$-wave and $f$-wave type similar to the AA ordering case we considered before. These observed odd-parity and $d$-wave (proportional to single $k_{\pm}^2$) symmetries can be confirmed by the analytic expressions in Eqs. (\ref{intraAB}). In Fig. \ref{fig2}(b), for the interlayer BCS coupling, the only component that possesses effective $d$-wave symmetry is $f_{11}^{\da\da}$. Checking through Eqs. (\ref{interAB}), we see that $f_{11}^{\da\da}$ expression contains single $k_{+}^2$ which produces the $d$-wave symmetry. The components $f_{11}^{\ua\ua},f_{12}^{\ua\da},f_{12}^{\da\ua}$ are of effective $s$-wave type while $f_{11}^{\ua\da},f_{12}^{\ua\ua},f_{11}^{\da\ua},f_{12}^{\da\da}$ show $p$-wave symmetry beside terms proportional to $k_+k_-^2k_+^2$ and $k_+^2k_-$. Mathematically, the reason that $f_{11}^{\da\da}$ gains $d$-wave symmetry is the specific combination of the interlayer coupling $\tilde{\bm T}$ which has off-diagonal terms, and the phonon-mediated electron-electron coupling $\Delta_\text{B}$ that is on the off-diagonal entries of Eq. (\ref{BCSH}).

Our prediction of the symmetry changes and generation of Pspin superconducting correlations can be confirmed experimentally. A relevant experiment that can reveal distinctive and direct evidence for the prediction of symmetry change is a high-resolution angular point-contact tunneling spectroscopy experiment. The angular tunneling spectroscopy experiment can determine the angular dependence of superconducting correlations before and after introducing the displacement and twist into the layers. Therefore, one should be able to distinguish between angle-independent and angle-dependent symmetry classes. Another method that might be utilized in investigating the change in the symmetry profile of superconducting correlations discussed above is nonlinear Meissner effect to image the symmetry profile of pairing correlations \cite{arxiv_nonlinearMSE}.

\section{Shifted and twisted bilayer graphene}\label{sec3}

To study twisted BLG, one can assume that the top and bottom layers are rotated mutually by an amount of $\theta/2$, one clockwise and the other counter clockwise. Therefore, the uncoupled single layers can be described by $H_{1,2}^\text{S}(\pm\theta/2,\theta_\text{k})=\Theta_z(\pm\theta/2)H^\text{S}(\theta_\text{k})\Theta_z^\dag(\pm\theta/2)$, where $\theta_\text{k}$ is the direction of the particles' momentum in each layer and $\Theta_z(\theta)$ is the rotation operator around the $z$ axis perpendicular to the graphene plane. In this case, the twist affects the coupling matrix as well so that one finds $\tilde{\bm T}({\theta,\mathbf{ r}},\textbf{k})$ from Eq. (\ref{tmatrix}), now with the phase factor $\exp (i\textbf{g} \cdot{\mathbf{ r}})$ in which $\textbf{g}_j=\Theta_z(+\theta/2)\textbf{G}_j - \Theta_z(-\theta/2)\textbf{G}_j$. 
For commensurate rotation angles, e.g., $\theta=21.787^\circ, 38.213^\circ$ one finds an effective $4\times 4$ Hamiltonian in the momentum space for the twisted BLG \cite{tbg1,tbg-2}. As the twisting angle becomes smaller, the superlattice cell gets larger. Note that for the noncommensurate angles, the bilayer system is inhomogeneous in real space and a very large basis set is needed to describe it \cite{tbg15,tbg16,tbg17,tbg18,tbg19,teor_tbg8,teor_tbg10,teor_tbg11,teor_tbg12,teor_tbg13,teor_tbg14}. Also, an AA-stacked bilayer can be transformed into an AB stacking order by keeping layer $1$ fixed, i.e. $\textbf{u}_1=0$, and shifting layer $2$ by $\textbf{u}_{2}= \epsilon({\bm a}_1 + {\bm a}_2)/3$ in Eq. (\ref{tmatrix}). One can easily show that this transformation changes AA order into AB, AB into AC, AC into AA when $\epsilon$ is equal to unity.

Nevertheless, in what follows, we consider a generic matrix with four different components for the coupling term, namely:
\begin{equation}\label{coulingT}
\tilde{\bm T}=\left(\begin{array}{cc}
t_{11} & t_{12}\\
t_{21} & t_{22}
\end{array}\right),
\end{equation}
and derive the anomalous Green's function, and finally, investigate a specific case of shifted BLG. Note that our analytic results below are applicable to twisted BLG where the low-energy physics can be described by the effective Hamiltonian approach. One simply needs to replace appropriate coupling terms (i.e., $t_{11}$, $t_{12}$, $t_{21}$, $t_{22}$), describing a twist.

As an illustrative example, we consider the pristine case,  $\mu=0$, and only include the intralayer coupling: $\Delta_\text{S}\neq 0$, $\Delta_\text{B}= 0$. The results for the anomalous Green's function are:
\begin{subequations}\label{eq:tbgtra}
\begin{eqnarray}
&&\Omega f_{11}^{\ua\ua}= 0,
\\
&&\Omega f_{11}^{\ua\da}=-\Delta_\text{S}  \left(\Delta_\text{S} ^2+k_- k_++t_{11}
   t_{22}-t_{12} t_{21}+\omega_n^2\right),
\\
&&\Omega f_{12}^{\ua\ua}=-\Delta_\text{S}  k_- (t_{22}+t_{22}^*)-i \Delta_\text{S}  \omega_n
   (t_{12}+t_{21}^*),
\\
&&\Omega f_{12}^{\ua\da}=-\Delta_\text{S}  (-k_- t_{21}+k_+ t_{21}^*-i \omega_n
   (t_{11}-t_{22}^*)),
\\
&&\Omega f_{11}^{\da\ua}=+\Delta_\text{S}  \left(\Delta_\text{S} ^2+k_- k_++t_{11}
   t_{22}-t_{12} t_{21}+\omega_n^2\right),
\\
&&\Omega f_{11}^{\da\da}=0,
\\
&&\Omega f_{12}^{\da\ua}=-\Delta_\text{S}  (-k_- t_{12}^*+k_+ t_{12}-i \omega_n
   (t_{11}^*-t_{22})),
\\
&&\Omega f_{12}^{\da\da}=+\Delta_\text{S}  k_+ (t_{11}+t_{11}^*)+i \Delta_\text{S}  \omega_n
   (t_{12}^*+t_{21}),
\end{eqnarray}
\end{subequations}
\begin{eqnarray}
   &&
   \Omega=k_-^2 \left(k_+^2-t_{12}^* t_{21}\right)+k_-
   \left(-k_+ \left(-2 \left(\Delta_\text{S} ^2+\omega_n^2\right)+ \right.\right. \nonumber\\ &&\left.\left. t_{11} t_{22}^*+t_{11}^* t_{22}\right)-i
   \omega_n (t_{12}^* (t_{11}+t_{22})+  t_{21}
   (t_{11}^*+t_{22}^*))\right)  \nonumber\\ 
   &&  -k_+^2 t_{12} t_{21}^*-i
   k_+ \omega_n (t_{21}^*
   (t_{11}+t_{22})+t_{12}
   (t_{11}^*+t_{22}^*))+ \nonumber\\ 
   &&\omega_n^2 \left(2 \Delta_\text{S}
   ^2+t_{11} t_{11}^*+t_{12} t_{12}^*+t_{21}
   t_{21}^*+    t_{22} t_{22}^*\right)-   \nonumber\\ 
   && \left(-\Delta_\text{S} ^2-t_{11}
   t_{22}+t_{12} t_{21}\right)   \left(\Delta_\text{S} ^2+t_{11}^*
   t_{22}^*-t_{12}^* t_{21}^*\right)+\omega_n^4.
\end{eqnarray}
An examination of these equations leads to the following conclusions. Similar to Sec. \ref{subsec21}, the generic coupling matrix leads to vanishing $f_{11}^{\ua\ua}, f_{11}^{\da\da}$. However, the symmetries are no longer similar to those explored in Sec. \ref{subsec21}. As seen, here, $\Omega$ includes terms of single $k_\pm^2$ that makes substantial changes. These terms have $d$-wave symmetry and thus the resultant Green's functions and corresponding superconducting correlations can now carry this symmetry as well. 

Next, we consider the interlayer coupling ($\Delta_S=0, \Delta_B\ne 0$ at $\mu=0$). The pairing correlations are given in Eqs. (\ref{eq:tbgter}):
\begin{subequations}\label{eq:tbgter}
\begin{eqnarray}
&&\Omega f_{11}^{\ua\ua}= 0,
\\
&&\Omega f_{11}^{\ua\da}=\Delta_\text{B}  (k_- t_{21}+k_+ t_{12}+i \omega_n
   (t_{11}+t_{22})),
\\
&&\Omega f_{12}^{\ua\ua}=i \Delta_\text{B}  (t_{21}^* t_{22}-t_{12} t_{22}^*),
\\
&&\Omega f_{12}^{\ua\da}=-\Delta_\text{B}  \left(\Delta_\text{B} ^2+k_- k_+-t_{11}
   t_{22}^*+t_{21} t_{21}^*+\omega_n^2\right),
\\
&&\Omega f_{11}^{\da\ua}=\Delta_\text{B}  (k_- t_{21}+k_+ t_{12}+i \omega_n
   (t_{11}+t_{22})),
\\
&&\Omega f_{11}^{\da\da}=0,
\\
&&\Omega f_{12}^{\da\ua}=\Delta_\text{B}  \left(\Delta_\text{B} ^2+k_- k_+-t_{11}^*
   t_{22}+t_{12} t_{12}^*+\omega_n^2\right),
\\
&&\Omega f_{12}^{\da\da}=-i \Delta_\text{B}  (t_{11} t_{12}^*-t_{11}^* t_{21}),
\\ &&
\Omega=-\Delta_\text{B} ^4-k_-^2 k_+^2+k_-^2 t_{12}^* t_{21}-2
   \Delta_\text{B} ^2 k_- k_+-  \nonumber \\ 
   &&  \omega_n^2 \left(2 \Delta_\text{B} ^2+ 2
   k_- k_++t_{11} t_{11}^*+t_{12}
   t_{12}^*+t_{21} t_{21}^*+t_{22} t_{22}^*\right)+   \nonumber \\ 
   &&  i
   \omega_n (k_- t_{12}^*
   (t_{11}+t_{22})+  k_- t_{21}
   (t_{11}^*+t_{22}^*)+   k_+ t_{21}^*
   (t_{11}+t_{22})+ \nonumber \\ 
   &&  k_+ t_{12}
   (t_{11}^*+t_{22}^*))+k_- k_+ t_{11}
   t_{22}^*+  k_- k_+   t_{11}^* t_{22}+k_+^2
   t_{12} t_{21}^*-  \nonumber \\ 
   &&  t_{11} t_{11}^* t_{22}
   t_{22}^*+ t_{11} t_{12}^* t_{21}^* t_{22}+\Delta_\text{B} ^2
   t_{11} t_{22}^*+ t_{11}^* t_{12} t_{21} t_{22}^*+\Delta_\text{B}
   ^2 t_{11}^* t_{22}\nonumber \\  
   && -\Delta_\text{B} ^2 t_{12} t_{12}^*- t_{12}
   t_{12}^* t_{21} t_{21}^*-\Delta_\text{B} ^2 t_{21}
   t_{21}^*-\omega_n^4.\nonumber 
\end{eqnarray}
\end{subequations}
We see that here, also $f_{11}^{\ua\ua}, f_{11}^{\da\da}$ are zero. Similarly to the previous case, $\Omega$ includes single $k_\pm^2$ terms, carrying $d$-wave symmetry. Note that these single $k_\pm^2$ terms are not vanishing here because all the entries of coupling matrix $\tilde{\bm T}$ are assumed nonzero with no specific symmetry. In all the previous cases, with specific stacking orders, $\Omega$ includes powers of $k_\pm k_\mp$ that carry $s$-wave symmetry.

\begin{figure*}[t]
\includegraphics[width=18.0cm,height=4.20cm]{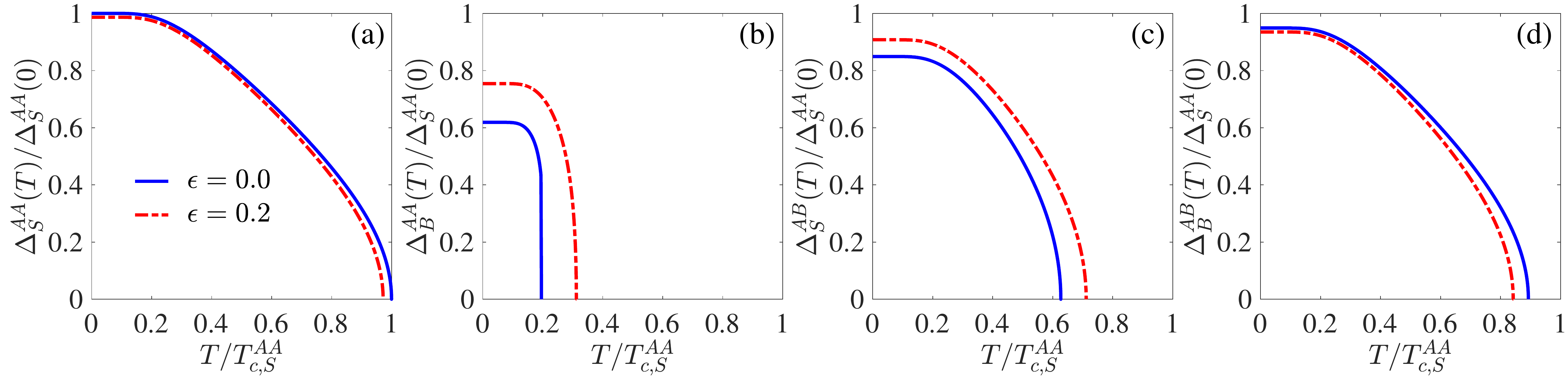}
\caption{(Color online).
Temperature dependence of intralayer $\Delta_\text{S}$ and interlayer $\Delta_\text{B}$ superconducting gaps for different stacking orders. The gap functions and temperature are normalized with respect to the intralayer gap at zero temperature for AA stacking without displacement, i.e., $\Delta_\text{S}^\text{AA}(T=0)$ and its critical temperature, i.e., $T^\text{AA}_{c,\text{S}}$, respectively.
 }\label{fig5}
\end{figure*}

At finite doping, the expressions become cumbersome, and we resort to numerical investigations. To examine the effect of displacement on the effective symmetry profile of pairing correlations, we study the shifted BLG scenario numerically. Figures \ref{fig3} and \ref{fig4} are shifted ($\epsilon=0.2$) counterparts of Figs. \ref{fig1} and \ref{fig2}. When comparing these figures, we see that, for AA stacking, the shift has induced imaginary parts into $f_{11}^{\ua\da}, f_{11}^{\da\ua}, f_{12}^{\ua\da}, f_{12}^{\da\ua}$ pairings with intralayer coupling and real parts for the interlayer coupling. Also, in the case of AB stacking, this displacement gives rise to nonzero real parts into $f_{12}^{\ua\da}, f_{12}^{\da\ua}$ with $\Delta_\text{S}\neq 0$ and nonzero imaginary parts into $f_{11}^{\ua\ua}, f_{12}^{\ua\da}$ with $\Delta_\text{B}\neq 0$. In all cases, the displacement tends to first destroy the symmetries and, subsequently, at large values of $\epsilon$ recovers either AA, AB, or AC symmetries depending on the initial ordering we start with. In the next section, we see how this deformation affects the critical temperature and superconducting gap and link them to the superconducting correlations discussed so far. 

As mentioned in passing, inspired by recent experiments in normal state\cite{herrero,C.Dean_exp}, BLG may develop a ``two-gap'' superconductivity if the electron-electron coupling strength within the layers is unequal to the electron-electron coupling between the two layers. Following this idea, we have calculated the anomalous Green's function when both $\Delta_\text{S}$ and $\Delta_\text{B}$ are nonzero. The results of this two-gap superconductivity scenario are given by Eqs. (\ref{eq:tbgtrater}) of the Appendix. In deriving  Eqs. (\ref{eq:tbgtrater}) we consider bulk properties , i.e., $\Delta=\Delta^\dag$ and assume that the two layers develop identical superconducting gap $\Delta_\text{S}$, which is unequal to the interlayer superconducting gap $\Delta_\text{B}$. To be specific, we consider twisted bilayers by an amount $\theta$ with zero doping, i.e., $\mu=0$. The intralayer Pspin-triplet components remains zero. Similarly to the single gap superconductivity case, $\Omega$ contains $k_\pm^2$ terms due to nonzero coupling potential terms considered.

\subsection{Temperature dependence of superconducting gaps }\label{subsec31}

Recalling Eqs. (\ref{intraAA})-(\ref{interAB}), the AB-stacked graphene bilayer with finite doping can potentially support a number of different pairings: $s$-wave, $p$-wave, $d$-wave, $f$-wave, and their combinations. In addition to the anomalous Green's function, an interaction potential $V({\textbf{k},\textbf{k}'})$ enters the equations, determining the superconducting gap function and critical temperature. In general, the superconducting gap function can be momentum dependent and it satisfies the following equation that should be solved self-consistently together with Eq. (\ref{BCSH}):
\begin{equation}\label{gapfunc}
\Delta^{\alpha\alpha'}_{\beta\beta'}({\textbf{k}})=-\sum_{\textbf{k}'} V^{\alpha\alpha',\sigma\sigma'}_{\beta\beta',\rho\rho'}({\textbf{k},\textbf{k}'}) f^{\sigma\sigma'}_{\rho\rho'}(\textbf{k}'),
\end{equation}
in which $\alpha, \alpha', \beta, \beta'$ are Pspin and layer indexes, respectively. The interaction potential can be both spin and layer dependent. Therefore, one can expand the interaction potential in spherical harmonics and in terms of the atomic orbital symbols introduced in Table \ref{table3}.

\begin {table}[b]
\caption {Phonon mediated electron-electron interaction potentials. } \label{table3}
\normalsize
\begin{center}
\begin{tabular}{|cc|}
\hline
\hline
\multicolumn{1}{|c|} {$V_s\propto 1$} & {$V_{\text{extended}-s}\propto k_x^2+k_y^2$} \\
\hline \multicolumn{2}{|c|} {$V_{p_x}\propto k_x$}   \\
 \hline\multicolumn{1}{|c|}{$V_{d_{xy}} \propto k_xk_y$} & {$V_{d_{x^2-y^2}}\propto k_x^2-k_y^2$}\\
  \hline\multicolumn{2}{|c|}  {$V_{f_{y(3x^2-y^2)}}\propto k_y(3k_x^2-k_y^2)$}  \\
   \hline\multicolumn{1}{|c|} {$V_{g_{xy(x^2-y^2)}}\propto $} & {$V_{g_{x^4+y^4}} \propto k_x^2(k_x^2-3k_y^2)$}\\
   \multicolumn{1}{|c|} {$k_xk_y(k_x^2-k_y^2)$}& \multicolumn{1}{|c|} {$- k_y^2(3k_x^2-k_y^2)$}\\
\hline
\hline
\end{tabular}
\end{center}
\end{table}
The type and details of phonon-mediated electron-electron interactions can be obtained through optical absorption and Raman spectroscopy experiments \cite{tbg2,tbg3,tbg4,tbg5,tbg6,tbg8,tbg9,tbg10,tbg11,tbg12,tbg13,tbg14,2ph-el_exp}. With the relevant anomalous Green's functions at hand, Eqs. (\ref{intraAA})-(\ref{eq:tbgtrater}), it is sufficient to insert the interaction potential into Eq. (\ref{gapfunc}) to obtain a self-consistent equation for the gap function, and solve it at different temperatures. As an example, we consider $s$-wave pairing (i.e., $V_s$ in Table \ref{table3}) and plot the superconducting gap as a function of temperature for different scenarios presented in Fig. \ref{fig5}. Note that we have used an identical interaction potential for intralayer and interlayer coupling potential in AA and AB orderings and set $\mu/t=1.5$ throughout our calculations (increasing the density of charged particles enhances the superconducting gap and critical temperature). With these assumptions, we find that the intralayer phonon-mediated electron-electron coupling in AA ordering hosts the largest superconducting gap that, from energy point of view, results in the lowest ground-state energy. We thus normalize all other superconducting gaps and temperature by the intralayer zero temperature superconducting gap of AA ordering, i.e., $\Delta_\text{S}^\text{AA}(T=0)$ and its critical temperature $T_{c,\text{S}}^\text{AA}$, respectively. Figure \ref{fig5} illustrates that the interlayer phonon-mediated electron-electron coupling in AA ordering results in the smallest critical temperature (Fig. \ref{fig5}(a)) while its AB ordering counterpart Fig. \ref{fig5}(d) has the closest gap amplitude and critical temperature to the intralayer AA ordering (the largest one). We have also applied a small displacement, i.e., $\epsilon=0.2$, according to the discussion in the previous subsection. Compairing Figs. \ref{fig5}(a) and \ref{fig5}(d), we find that the intralayer and interlayer couplings in AA and AB stacking orders share similar behavior in all features in the presence and absence of the displacement. While the displacement in intralayer coupling of AA stacking, Fig. \ref{fig5}(a), decreases the gap amplitude and critical temperature, it enhances these quantities in AB ordering, Fig. \ref{fig5}(c). This finding is reversed for the interlayer BCS coupling scenario in AA configuration. The real parts of $f_{11}^{\ua\da}, f_{11}^{\da\ua}$ and $f_{12}^{\ua\da}, f_{12}^{\da\ua}$ play a crucial role resulting in dramatic changes to the gap function and critical temperature. The underlying reason for these changes can be understood by compairing Figs. \ref{fig1} and \ref{fig2} with Figs. \ref{fig3} and \ref{fig4}. The displacement $\epsilon$ induces real parts to $f_{11}^{\ua\da}$ and $f_{12}^{\ua\da}$ with interlayer and intralyer phonon-mediated electron-electron coupling of AA and AB orderings, respectively, Figs. \ref{fig5}(b) and \ref{fig5}(c). However, a nonzero displacement $\epsilon$ applied to AA and AB orderings with intralayer and interlayer BCS coupling, respectively, induces nonzero imaginary parts to $f_{11}^{\ua\da}$ and $f_{12}^{\ua\da}$ that changes slightly the gap function and critical temperature, Figs. \ref{fig5}(a) and \ref{fig5}(d).

One can repeat the study presented in Fig. \ref{fig5} for other interaction types given in Table \ref{table3} and the given Green's functions in Eqs. (\ref{intraAA})-(\ref{eq:tbgtrater}). Nevertheless, we postpone such a study to future works when more details of interactions or relevant experimental data are available. It is, however, apparent that considering $d$-wave or $g$-wave interactions in Table \ref{table3} for the electron-electron coupling potential $V({\textbf{k},\textbf{k}'})$ with its spin state being singlet, the ratio of superconducting gap and critical temperature increases (according to the traditional calculations of such interaction potentials \cite{mineev}) for cases where the Green's function itself possesses $d$-wave symmetry, as discussed in the previous section. Therefore, such a discussion should be relevant for the recent experiment \cite{herrero} that observed a puzzling behavior combining the features of high $T_c$ and conventional superconductors. Accordingly, by deforming the BLG, $d$-wave spin-singlet pairing may become dominant and increase the ratio of the superconducting gap and $T_c$.

In our calculations above, we have simply followed the BCS picture of superconductivity where the coupling of two particles with opposite spins and momenta are building blocks of superconductivity. Additionally, we have assumed that the influence of twist and displacement in superconducting BLG are encoded in the coupling terms of Eq.~(\ref{coulingT}) and the electron-electron coupling interaction potentials remain unchanged. Nevertheless, if twist and displacement change the amplitude of electron-electron coupling interaction potentials only, the above findings remain intact. However, if twist and displacement introduce momentum-dependent changes to the electron-electron coupling interaction potentials, similar to those presented in Table \ref{table3}, one should repeat the above calculations with these momentum-dependent potentials.

\section{summary}\label{sec4}

In summary, using an effective Hamiltonian approach to discuss in-plane displacements between graphene monolayers in a bilayer graphene system, we investigate the superconducting pairing correlations by deriving the anomalous Green's function. Motivated by a recent experiment, we consider both intralayer and interlayer phonon-mediated electron-electron couplings in AA and AB stacking orders. Our results reveal that both AA and AB configurations at the charge neutrality point, $\mu=0$, can only develop even-parity $s$-wave and odd-parity $p$-wave superconducting correlations. At a finite doping, $\mu\neq 0$, this finding remains intact in a AA system with the addition of odd-parity $f$-wave symmetry while AB ordering, exclusively, can host even-parity $d$-wave symmetries. Introducing a generic coupling potential between graphene monolayers in bilayer graphene, we show that displacement of graphene monolayers can induce $d$-wave symmetry at $\mu=0$ as well. Our results suggest that a switching to $d$-wave symmetry can be achieved by a simple displacement of the two coupled pristine graphene layers at a finite density of charged carriers. We also discuss the possible appearance of pseudospin-triplet and odd-frequency pairings. Finally, we consider an $s$-wave interaction potential and study the superconducting gap function and critical temperature in a specific scenario of displacement (in-plane shifting of the monolayers of graphene) in AA and AB stacking orders. We find that AA stacking with an intralayer electron-electron coupling most favorably hosts $s$-wave superconductivity while AA ordering with interlayer electron-electron coupling is the next desirable platform. Also, the exertion of a slight in-plane shift in bilayer graphene can increase both the amplitude of the superconducting gap and critical temperature in AA and AB orderings with phonon-mediated interlayer and intralayer electron-electron couplings.

\begin{acknowledgments}
M.A. is supported by Iran's National Elites Foundation (INEF). Center for Nanostructured Graphene is supported by the Danish National Research Foundation (Project No. DNRF103).
\end{acknowledgments}

\onecolumngrid
\appendix

\section{Hamiltonian of SLG with superconductivity}\label{apdx1}

A monolayer graphene can be described by two valleys in $k$-space: {\bf K} =$\frac{2\pi}{3a}(1,+1/\sqrt{3})$ and {\bf K}$'$=$\frac{2\pi}{3a}(1,-1/\sqrt{3})$. At low energies, the dispersion relation near the corners of the Brillouin zone, i.e., {\bf K} and {\bf K}$'$ are linear and can be described by a Dirac Hamiltonian: $H_{\bf K}(\textbf{k}) = \hbar v_\textbf{F}\textbf{k}\cdot {\bm \sigma}$.
The corresponding wavefunction is a two-component spinor, namely,
\begin{equation}
\hbar v_\textbf{F}\left(\begin{array}{cc}
0 & k_x-ik_y\\
k_x+ik_y & 0
\end{array}\right )\left(\begin{array}{c}
\psi_{\text{A}{\bf K}}(k)\\
\psi_{\text{B}{\bf K}}(k)
\end{array}\right )=\varepsilon \left(\begin{array}{c}
\psi_{\text{A}{\bf K}}(k)\\
\psi_{\text{B}{\bf K}}(k)
\end{array}\right ).
\end{equation} 
Here, A and B denote the sublattices of graphene honeycomb lattice. Note that the Hamiltonian above couples these two sublattices. This appears because the nearest neighbor site of each sublattice site belongs to the other sublattice. This can be clearly seen in Fig. \ref{model}(a), where sublattices are marked by blue and red colors. Hence, the total spin-less Hamiltonian can be expressed by:
\begin{equation}\label{A2}
H(\textbf{k}) = \left(\begin{array}{cc}
H_{\bf K}(\textbf{k}) & 0\\
0& H_{{\bf K}'}(\textbf{k})
\end{array}\right ),
\end{equation} 
where $H_{{\bf K}'}(\textbf{k}) = \hbar v_\textbf{F}\textbf{k}\cdot {\bm \sigma}^*$. Next, we can add spin to this Hamiltonian by introducing Pauli matrices acting in spin-space: ${\bm \tau}=(\tau_x,\tau_y,\tau_z)$. In this case, the spin-full Hamiltonian can be expressed by $H_{tot}=H(\textbf{k})\tau_0$ where $\tau_0$ reflects the absence of spin coupling. In SLG the two valleys are decoupled in the absence of an external potential\cite{mm}. Note that the form Eq. (\ref{A2}) applies even in the presence of elastic impurities and disorder. Therefore, the Hamiltonian becomes degenerated and reduces to a $4\times 4$ Hamiltonian. Now we incorporate superconductivity by introducing a two-electron amplitude,
\begin{equation}\label{Ds}
\Delta_\text{S} \Big\langle\psi^\dag_{A\uparrow {\bf K}}\psi^\dag_{B\downarrow {\bf K}'}\Big\rangle+\text{H.c.},
\end{equation}
in which $\Delta_\text{S}$ is the gap representing BCS spin-singlet phonon-mediated electron-electron coupling between sublattices A, B and valleys {\bf K}, {\bf K}$'$. Note that there are several options to introduce a two-electron amplitude in the presence of superconductivity, including Pspin-triplet coupling that couples two electrons in the same sublattice \cite{BP1-prb2018,BP2-prb2018,BP2019}. In the main text, we have used the above two-electron amplitude Eq. (\ref{Ds}) in our calculations. It is worth mentioning that the hole-excitation block of the Hamiltonian in the particle-hole space, when introducing superconductivity, is equivalent to the Hamiltonian of {\bf K}$'$ valley. Therefore, superconductivity can be viewed as a mean to introduce valley coupling in SLG. When two normal SLGs are coupled (making a BLG), more options for incorporating the two-electron amplitudes in its superconducting phase are generated. To clarify the presence of the second SLG, we have introduced indices $1,2$ that label top and bottom SLGs as described in the main text. In this case, to be consistent with the previous case, we consider the following two-electron amplitude:   
\begin{equation}
\Delta_\text{B} \Big\langle\psi^\dag_{1A\uparrow {\bf K}}\psi^\dag_{2B\downarrow {\bf K}'}\Big\rangle+\text{H.c.},
\end{equation}
where $\Delta_\text{B}$ is the gap representing BCS spin-singlet phonon-mediated electron-electron coupling between sublattices A, B, valleys {\bf K}, {\bf K}$'$, and layers 1,2. To simplify our notation throughout the presentation in the main text, due to the spin-degeneracy\cite{RMP-2008-Beenakker, RMP-2009-Neto,PRL-2013-Halter}, we have dropped spin and valley {\bf K}, {\bf K}$'$ indices and only keep the sublattice indices. Since we discuss AA and AB stacking orders, from now on, we change our notation of A,B sublattices to $\uparrow$,$\downarrow$ and call them pseudospin (Pspin).

 \section{Green's function of displaced BLG with superconductivity}\label{apdx2}
In this part, we present expressions derived for the anomalous Green's function in the presence of a finite doping, considering intralayer and interlayer phonon-mediated electron-electron couplings.
\\*
\\*
$\bullet$\textbf{Intralayer BCS coupling, $\Delta_\text{S}\neq 0, \Delta_\text{B}= 0$, in AA stacking order}
\begin{subequations}\label{intraAA}
\begin{eqnarray}
&\Omega f_{11}^{\ua\ua}= +2 \Delta_\text{S}  k_- \mu  \left(\left(\Delta_\text{S} ^2+\mu ^2+\omega_n^2\right)^2+k_-^2 k_+^2+2 k_- k_+ \left(\Delta_\text{S} ^2-\mu
   ^2+t^2+\omega_n^2\right)-  3 t^4-2 t^2 \left(\Delta_\text{S} ^2-\mu ^2+\omega_n^2\right)\right),
\\
&\Omega f_{11}^{\ua\da}=-\Delta_\text{S}  \left(k_-^3 k_+^3-k_-^2 k_+^2 \left(-3 \Delta_\text{S} ^2+\mu ^2+t^2-3
   \omega_n^2\right)+k_- k_+ \left(\left(\Delta_\text{S} ^2+\mu ^2+\omega_n^2\right) \left(3 \Delta_\text{S} ^2-\mu ^2+3 \omega_n^2\right)-t^4+2 t^2 \left(\Delta_\text{S}
   ^2+5 \mu ^2+\omega_n^2\right)\right)+ \right. \nonumber\\ &\left. \left(\Delta_\text{S} ^2+(t-\mu )^2+\omega_n^2\right) \left(\Delta_\text{S} ^2+\mu ^2+t^2+\omega_n^2\right) \left(\Delta_\text{S} ^2+(\mu
   +t)^2+\omega_n^2\right)\right),
\\&
\Omega f_{12}^{\ua\ua}=-2 \Delta_\text{S}  k_- t \left(\left(\Delta_\text{S} ^2-3 \mu ^2+\omega_n^2\right) \left(\Delta_\text{S}
   ^2+\mu ^2+\omega_n^2\right)+k_-^2 k_+^2+2 k_- k_+
   \left(\Delta_\text{S} ^2+\mu ^2-t^2+\omega_n^2\right)+t^4+2 t^2 \left(\Delta_\text{S} ^2+\mu
   ^2+\omega_n^2\right)\right),
\\&
\Omega f_{12}^{\ua\da}=-2 \Delta_\text{S}  \mu  t \left(-3 k_-^2 k_+^2+2 k_- k_+ \left(-\Delta_\text{S} ^2+\mu
   ^2+t^2-\omega_n^2\right)+\left(\Delta_\text{S} ^2+(t-\mu )^2+\omega_n^2\right)
   \left(\Delta_\text{S} ^2+(\mu +t)^2+\omega_n^2\right)\right),
\\&
\Omega f_{11}^{\da\ua}=-\Delta_\text{S}  \left(-k_-^3 k_+^3+k_-^2 k_+^2 \left(-3 \Delta_\text{S} ^2+\mu ^2+t^2-3
   \omega_n^2\right)+k_- k_+ \left(-\left(\Delta_\text{S} ^2+\mu ^2+\omega_n^2\right) \left(3 \Delta_\text{S} ^2-\mu ^2+3 \omega_n^2\right)+t^4-2 t^2 \left(\Delta_\text{S}
   ^2+5 \mu ^2+\omega_n^2\right)\right)-  \right. \nonumber\\ &\left.  \left(\Delta_\text{S} ^2+(t-\mu )^2+\omega_n^2\right) \left(\Delta_\text{S} ^2+\mu ^2+t^2+\omega_n^2\right) \left(\Delta_\text{S} ^2+(\mu
   +t)^2+\omega_n^2\right)\right),
\\&
\Omega f_{11}^{\da\da}=-2 \Delta_\text{S}  k_+ \mu  \left(\left(\Delta_\text{S} ^2+\mu ^2+\omega_n^2\right)^2+k_-^2
   k_+^2+2 k_- k_+ \left(\Delta_\text{S} ^2-\mu ^2+t^2+\omega_n^2\right)-3
   t^4-2 t^2 \left(\Delta_\text{S} ^2-\mu ^2+\omega_n^2\right)\right),
\\&
\Omega f_{12}^{\da\ua}=+2 \Delta_\text{S}  \mu  t \left(-3 k_-^2 k_+^2+2 k_- k_+ \left(-\Delta_\text{S} ^2+\mu
   ^2+t^2-\omega_n^2\right)+\left(\Delta_\text{S} ^2+(t-\mu )^2+\omega_n^2\right)
   \left(\Delta_\text{S} ^2+(\mu +t)^2+\omega_n^2\right)\right),
\\&
\Omega f_{12}^{\da\da}=+2 \Delta_\text{S}  k_+ t \left(\left(\Delta_\text{S} ^2-3 \mu ^2+\omega_n^2\right) \left(\Delta_\text{S}
   ^2+\mu ^2+\omega_n^2\right)+k_-^2 k_+^2+2 k_- k_+
   \left(\Delta_\text{S} ^2+\mu ^2-t^2+\omega_n^2\right)+t^4+2 t^2 \left(\Delta_\text{S} ^2+\mu
   ^2+\omega_n^2\right)\right),\\&
   \Omega=\left(k_-^2 k_+^2+2 k_- k_+ \left(\Delta_\text{S} ^2-(t-\mu )^2+\omega_n^2\right)+\left(\Delta_\text{S} ^2+(t-\mu )^2+\omega_n^2\right)^2\right)
   \left(k_-^2 k_+^2-2 k_- k_+ \left(-\Delta_\text{S} ^2+(\mu
   +t)^2-\omega_n^2\right)+\left(\Delta_\text{S} ^2+(\mu +t)^2+\omega_n^2\right)^2\right),\;\;\;\;\;
\end{eqnarray}
\end{subequations}
\\*
$\bullet$\textbf{Intralayer BCS coupling, $\Delta_\text{S}\neq 0, \Delta_\text{B}= 0$, in AB stacking order}
\begin{subequations}\label{intraAB}
\begin{eqnarray}
&\Omega f_{11}^{\ua\ua}= +2 \Delta_\text{S}  \mu  \left(k_-^3 k_+^2+2 k_-^2 k_+ \left(\Delta_\text{S} ^2-\mu
   ^2+\omega_n^2\right)+k_- \left(\Delta_\text{S} ^2+\mu ^2+\omega_n^2\right)^2-k_+ t^2 \left(\mu ^2+\omega_n^2\right)\right),
\\
&\Omega f_{11}^{\ua\da}=-\Delta_\text{S}  \left(k_-^3 k_+^3+k_-^2 k_+^2 \left(3 \Delta_\text{S} ^2-\mu ^2+3
   \omega_n^2\right)+k_- k_+ \left(\left(\Delta_\text{S} ^2+\mu ^2+\omega_n^2\right) \left(3 \Delta_\text{S} ^2-\mu ^2+3 \omega_n^2\right)+t^2 \left(\mu
   ^2+\omega_n^2\right)\right)-   2 i \mu  t^2 \omega_n \left(\Delta_\text{S}
   ^2+k_+^2+\mu ^2\right)-\right.\nonumber\\ & \left. 2 k_+^2 \mu ^2 t^2+\omega_n^4 \left(3
   \left(\Delta_\text{S} ^2+\mu ^2\right)+t^2\right)+\omega_n^2 \left(3 \left(\Delta_\text{S} ^2+\mu
   ^2\right)^2+\Delta_\text{S} ^2 t^2\right)-   2 i \mu  t^2 \omega_n^3+\left(\Delta_\text{S} ^2+\mu
   ^2\right) \left(\Delta_\text{S} ^2+\mu  (\mu -t)\right) \left(\Delta_\text{S} ^2+\mu  (\mu
   +t)\right)+\omega_n^6\right),
\\
&\Omega f_{12}^{\ua\ua}=-\Delta_\text{S}  t \left(-2 k_-^3 k_+ \mu +k_-^2 \left(2 \mu  \left(-\Delta_\text{S} ^2+(\mu
   +i \omega_n)^2\right)+k_+^2 (\mu +i \omega_n)\right)+2 i k_-
   k_+   \omega_n \left(\Delta_\text{S} ^2+\mu ^2+\omega_n^2\right)- \right.\nonumber\\ & \left. (\mu -i
   \omega_n) \left(\Delta_\text{S} ^2+\mu ^2-t (\mu +i \omega_n)+\omega_n^2\right) \left(\Delta_\text{S} ^2+\mu  (\mu +t)+i t \omega_n+\omega_n^2\right)\right),
\\
&\Omega f_{12}^{\ua\da}=+2 \Delta_\text{S}  \mu  t \left(-k_-^2 k_+ (\mu -i \omega_n)+k_- (\mu +i
   \omega_n) \left(\Delta_\text{S} ^2+k_+^2+\mu ^2+\omega_n^2\right)-k_+
   (\mu -i \omega_n) \left(\Delta_\text{S} ^2+\mu ^2+\omega_n^2\right)\right),
\\
&\Omega f_{11}^{\da\ua} =-\Delta_\text{S}  \left(-k_-^3 k_+^3+k_-^2 k_+^2 \left(-3 \Delta_\text{S} ^2+\mu ^2-3
   \omega_n^2\right)-k_- k_+ \left(\left(\Delta_\text{S} ^2+\mu ^2+\omega_n^2\right) \left(3 \Delta_\text{S} ^2-\mu ^2+3 \omega_n^2\right)+t^2 \left(\mu
   ^2+\omega_n^2\right)\right)-   2 i \mu  t^2 \omega_n \left(\Delta_\text{S}
   ^2+k_+^2+\mu ^2\right)\right.\nonumber\\ &\left. +2 k_+^2 \mu ^2 t^2-\omega_n^4 \left(3
   \left(\Delta_\text{S} ^2+\mu ^2\right)+t^2\right)-\omega_n^2 \left(3 \left(\Delta_\text{S} ^2+\mu
   ^2\right)^2+\Delta_\text{S} ^2 t^2\right)-   2 i \mu  t^2 \omega_n^3-\left(\Delta_\text{S} ^2+\mu
   ^2\right) \left(\Delta_\text{S} ^2+\mu  (\mu -t)\right) \left(\Delta_\text{S} ^2+\mu  (\mu
   +t)\right)-\omega_n^6\right),
\\
&\Omega f_{11}^{\da\da}=-2 \Delta_\text{S}  \mu  \left(k_-^2 k_+^3+k_- k_+^2 \left(2 \left(\Delta_\text{S} ^2-\mu
   ^2+\omega_n^2\right)+t^2\right)+\Delta_\text{S} ^2 k_- t^2-k_+^3 t^2-k_+
   \left(\Delta_\text{S} ^2+\mu ^2+\omega_n^2\right) \left(-\Delta_\text{S} ^2-\mu ^2+t^2-\omega_n^2\right)\right),
\\
&\Omega f_{12}^{\da\ua}=-\Delta_\text{S}  t \left(2 i \mu  \omega_n (k_-+k_+) \left(\Delta_\text{S} ^2+k_-
   k_++\mu ^2\right)+2 i \mu  \omega_n^3
   (k_-+k_+)+(k_+-k_-) \left(\left(\Delta_\text{S} ^2+k_-
   k_+\right)^2  -\mu ^4+\mu ^2 t^2\right)- \right.\nonumber\\ &\left.  \omega_n^2 (k_--k_+)
   \left(2 \Delta_\text{S} ^2+2 k_- k_++t^2\right)+\omega_n^4
   (k_+-k_-)\right),
\\
&\Omega f_{12}^{\da\da}=-\Delta_\text{S}  t \left(k_-^2 k_+^2 (\mu -i \omega_n)-2 k_- k_+
   \left(k_+^2 \mu +i \omega_n \left(\Delta_\text{S} ^2+\mu ^2+\omega_n^2\right)\right)-2 k_+^2 \mu  \left(\Delta_\text{S} ^2-(\mu -i \omega_n)^2\right)- \right.\nonumber\\ &\left.  (\mu +i \omega_n) \left(\Delta_\text{S} ^2+\mu  (\mu +t)-i t \omega_n+\omega_n^2\right) \left(\Delta_\text{S} ^2+\mu ^2-\mu  t+i t
   \omega_n+\omega_n^2\right)\right),\\
   &\Omega=\Delta_\text{S} ^8+4 \Delta_\text{S} ^6 \left(\mu ^2+\omega_n^2\right)+k_-^4 k_+^4+4
   k_-^3 k_+^3 \left(\Delta_\text{S} ^2-\mu ^2+\omega_n^2\right)+2 k_-^2
   \left(k_+^2 \left(3 \Delta_\text{S} ^4-2 \Delta_\text{S} ^2 \mu ^2+3 \mu ^4+\omega_n^2 \left(6
   \Delta_\text{S} ^2-2 \mu ^2+t^2\right)-\mu ^2 t^2+3 \omega_n^4\right) \right. \nonumber \\& \left. -2 \Delta_\text{S} ^2 \mu ^2
   t^2\right)+4 k_- k_+ \left(\Delta_\text{S} ^2+\mu ^2+\omega_n^2\right)
   \left(\left(\Delta_\text{S} ^2+\omega_n^2\right)^2-\mu ^4+t^2 \left(\mu ^2+\omega_n^2\right)\right)+4 \Delta_\text{S} ^2 \left(-k_+^2 \mu ^2 t^2+\left(\mu ^2+\omega_n^2\right)^3+t^2 \left(\omega_n^4-\mu ^4\right)\right)  \nonumber \\& +2 \Delta_\text{S} ^4 \left(3
   \left(\mu ^2+\omega_n^2\right)^2+t^2 \left(\omega_n^2-\mu
   ^2\right)\right)+\left(\mu ^2+\omega_n^2\right)^2 \left((t-\mu )^2+\omega_n^2\right) \left((\mu +t)^2+\omega_n^2\right),
\end{eqnarray}
\end{subequations}
\\*
$\bullet$\textbf{Interlayer BCS coupling, $\Delta_\text{S}= 0, \Delta_\text{B}\neq 0$, in AA stacking order}
\begin{subequations}\label{interAA}
\begin{eqnarray}
&\Omega f_{11}^{\ua\ua}= 8 i \Delta_\text{B}  k_- \mu  t \omega_n \left(\Delta_\text{B} ^2+k_- k_++\mu
   ^2-t^2+\omega_n^2\right),
\\
&\Omega f_{11}^{\ua\da}=2 i \Delta_\text{B}  t \omega_n \left(k_-^2 k_+^2+2 k_- k_+
   \left(\Delta_\text{B} ^2+3 \mu ^2-t^2+\omega_n^2\right)+2 \omega_n^2 \left(\Delta_\text{B}
   ^2+\mu ^2+t^2\right)+\left(\Delta_\text{B} ^2+\mu ^2-t^2\right)^2+\omega_n^4\right),
\\
&\Omega f_{12}^{\ua\ua}=2 \Delta_\text{B}  k_- \mu  \left(\left(\Delta_\text{B} ^2+\mu ^2+\omega_n^2\right)^2+k_-^2
   k_+^2+2 k_- k_+ \left(\Delta_\text{B} ^2-\mu ^2-t^2+\omega_n^2\right)+t^4-2 t^2 \left(\Delta_\text{B} ^2+\mu ^2+3 \omega_n^2\right)\right),
\\
&\Omega f_{12}^{\ua\da}=-\Delta_\text{B}  \left(\Delta_\text{B} ^2+k_- k_++\mu ^2-t^2+\omega_n^2\right)
   \left(k_-^2 k_+^2+2 k_- k_+ \left(\Delta_\text{B} ^2-\mu
   ^2-t^2+\omega_n^2\right)+2 \omega_n^2 \left(\Delta_\text{B} ^2+\mu
   ^2+t^2\right)+\left(\Delta_\text{B} ^2+\mu ^2-t^2\right)^2+\omega_n^4\right),
\\
&\Omega f_{11}^{\da\ua}=2 i \Delta_\text{B}  t \omega_n \left(k_-^2 k_+^2+2 k_- k_+
   \left(\Delta_\text{B} ^2+3 \mu ^2-t^2+\omega_n^2\right)+2 \omega_n^2 \left(\Delta_\text{B}
   ^2+\mu ^2+t^2\right)+\left(\Delta_\text{B} ^2+\mu ^2-t^2\right)^2+\omega_n^4\right),
\\
&\Omega f_{11}^{\da\da}=-8 i \Delta_\text{B}  k_+ \mu  t \omega_n \left(\Delta_\text{B} ^2+k_- k_++\mu
   ^2-t^2+\omega_n^2\right),
\\
&\Omega f_{12}^{\da\ua}=\Delta_\text{B}  \left(\Delta_\text{B} ^2+k_- k_++\mu ^2-t^2+\omega_n^2\right)
   \left(k_-^2 k_+^2+2 k_- k_+ \left(\Delta_\text{B} ^2-\mu
   ^2-t^2+\omega_n^2\right)+2 \omega_n^2 \left(\Delta_\text{B} ^2+\mu
   ^2+t^2\right)+\left(\Delta_\text{B} ^2+\mu ^2-t^2\right)^2+\omega_n^4\right),
\\
&\Omega f_{12}^{\da\da}=-2 \Delta_\text{B}  k_+ \mu  \left(\left(\Delta_\text{B} ^2+\mu ^2+\omega_n^2\right)^2+k_-^2 k_+^2+2 k_- k_+ \left(\Delta_\text{B} ^2-\mu
   ^2-t^2+\omega_n^2\right)+t^4-2 t^2 \left(\Delta_\text{B} ^2+\mu ^2+3 \omega_n^2\right)\right),
   \\
   &\Omega=k_-^4 k_+^4+4 k_-^3 k_+^3 \left(\Delta_\text{B} ^2-\mu ^2-t^2+\omega_n^2\right)+2 k_-^2 k_+^2 \left(3 \Delta_\text{B} ^4-2 \Delta_\text{B} ^2 \mu ^2+3 \mu ^4+3
   t^4-2 \omega_n^2 \left(-3 \Delta_\text{B} ^2+\mu ^2+t^2\right)-6 \Delta_\text{B} ^2 t^2+2 \mu ^2
   t^2+3 \omega_n^4\right)\nonumber\\&+4 k_- k_+ \left(\left(\Delta_\text{B} ^2-\mu
   ^2+\omega_n^2\right) \left(\Delta_\text{B} ^2+\mu ^2+\omega_n^2\right)^2-t^6+t^4
   \left(3 \Delta_\text{B} ^2+\mu ^2-\omega_n^2\right)+t^2 \left(-3 \Delta_\text{B} ^4-2 \Delta_\text{B} ^2
   \left(\mu ^2+\omega_n^2\right)+\right.\right.\nonumber\\&\left.\left.\mu ^4+10 \mu ^2 \omega_n^2+\omega_n^4\right)\right)+\left(2 \omega_n^2 \left(\Delta_\text{B} ^2+\mu
   ^2+t^2\right)+\left(\Delta_\text{B} ^2+\mu ^2-t^2\right)^2+\omega_n^4\right)^2,
\end{eqnarray}
\end{subequations}
\\*
$\bullet$\textbf{Interlayer BCS coupling, $\Delta_\text{S}= 0, \Delta_\text{B}\neq 0$, in AB stacking order}
\begin{subequations}\label{interAB}
\begin{eqnarray}
&\Omega f_{11}^{\ua\ua}= -2 \Delta_\text{B}  \mu  t \left(\left(\Delta_\text{B} ^2+\mu ^2+\omega_n^2\right)^2-k_-^2
   k_+^2\right),
\\
&\Omega f_{11}^{\ua\da}=-\Delta_\text{B}  k_+ t \left(k_-^2 k_+^2+2 k_- k_+ \left(\Delta_\text{B} ^2+\mu
   ^2+\omega_n^2\right)+\left(\Delta_\text{B} ^2+\mu ^2+\omega_n^2\right)
   \left(\Delta_\text{B} ^2-3 \mu ^2+t^2+\omega_n^2\right)\right),
\\
&\Omega f_{12}^{\ua\ua}=2 \Delta_\text{B}  k_- \mu  \left(\left(\Delta_\text{B} ^2+\mu ^2+\omega_n^2\right)^2+k_-^2
   k_+^2+k_- k_+ \left(2 \left(\Delta_\text{B} ^2-\mu ^2+\omega_n^2\right)+t^2\right)\right),
\\
&\Omega f_{12}^{\ua\da}=-\Delta_\text{B}  \left(\Delta_\text{B} ^2+k_- k_++\mu ^2+\omega_n^2\right)
   \left(k_-^2 k_+^2+2 k_- k_+ \left(\Delta_\text{B} ^2-\mu ^2+\omega_n^2\right)+\left(\Delta_\text{B} ^2+\mu ^2+\omega_n^2\right) \left(\Delta_\text{B} ^2+\mu
   ^2+t^2+\omega_n^2\right)\right),
\\
 &\Omega f_{11}^{\da\ua}=\Delta_\text{B}  k_+ t \left(k_-^2 k_+^2+2 k_- k_+ \left(\Delta_\text{B} ^2+\mu
   ^2+\omega_n^2\right)+\left(\Delta_\text{B} ^2+\mu ^2+\omega_n^2\right)
   \left(\Delta_\text{B} ^2-3 \mu ^2+t^2+\omega_n^2\right)\right),
\\
&\Omega f_{11}^{\da\da}=-2 \Delta_\text{B}  k_+^2 \mu  t \left(2 \left(\Delta_\text{B} ^2-\mu ^2+\omega_n^2\right)+2
   k_- k_++t^2\right),
\\
&\Omega f_{12}^{\da\ua}=\Delta_\text{B}  \left(k_-^3 k_+^3+k_-^2 k_+^2 \left(3 \Delta_\text{B} ^2-\mu ^2+t^2+3
   \omega_n^2\right)+k_- k_+ \left(\Delta_\text{B} ^2+\mu ^2+\omega_n^2\right) \left(3 \Delta_\text{B} ^2-\mu ^2+3 t^2+ 3 \omega_n^2\right)+\right.\nonumber \\&\left. \left(\Delta_\text{B}
   ^2+\mu ^2+\omega_n^2\right) \left(\Delta_\text{B} ^2+(t-\mu )^2+\omega_n^2\right)
   \left(\Delta_\text{B} ^2+(\mu +t)^2+\omega_n^2\right)\right),
\\
&\Omega f_{12}^{\da\da}=-2 \Delta_\text{B}  k_+ \mu  \left(\left(\Delta_\text{B} ^2+\mu ^2+\omega_n^2\right)^2+k_-^2 k_+^2+k_- k_+ \left(2 \left(\Delta_\text{B} ^2-\mu
   ^2+\omega_n^2\right)+t^2\right)\right),\\&
   \Omega=k_-^4 k_+^4+4 k_-^3 k_+^3 \left(\Delta_\text{B} ^2-\mu ^2+\omega_n^2\right)+2 k_-^2 k_+^2 \left(3 \Delta_\text{B} ^4-2 \Delta_\text{B} ^2 \left(\mu ^2-3
   \omega_n^2\right)+3 \mu ^4-2 \mu ^2 \omega_n^2+t^2 \left(\Delta_\text{B} ^2-\mu
   ^2+\omega_n^2\right)+3 \omega_n^4\right)+\nonumber \\& 4 k_- k_+
   \left(\Delta_\text{B} ^2+\mu ^2+\omega_n^2\right)^2 \left(\Delta_\text{B} ^2-\mu
   ^2+t^2+\omega_n^2\right)+\left(\Delta_\text{B} ^2+\mu ^2+\omega_n^2\right)^2
   \left(\Delta_\text{B} ^2+(t-\mu )^2+\omega_n^2\right) \left(\Delta_\text{B} ^2+(\mu
   +t)^2+\omega_n^2\right),
\end{eqnarray}
\end{subequations}
\\*
$\bullet$\textbf{Intralayer and interlayer BCS coupling, $\Delta_\text{S}\neq 0, \Delta_\text{B}\neq 0$, with a generic coupling matrix}
\begin{subequations}\label{eq:tbgtrater}
\begin{eqnarray}
&\Omega f_{11}^{\ua\ua}= 0,
\\
&\Omega f_{11}^{\ua\da}=-\Delta_\text{B}^2 \Delta_\text{S}-\Delta_\text{B} (k_-^- t_{21}+k_+^- t_{12}+i
   \omega_n (t_{11}+t_{22}))-  \Delta_\text{S} \left(\Delta_\text{S}^2+k_-^-
   k_+^-+t_{11} t_{22}-t_{12} t_{21}+\omega_n^2\right),
\\
&\Omega f_{12}^{\ua\ua}=-(\Delta_\text{S} k_-^+ t_{22} - \Delta_\text{B} t_{21}^* t_{22} + \Delta_\text{S} k_-^- t_{22}^* + \Delta_\text{B} t_{12} t_{22}^* +
 i (\Delta_\text{B} (-k_-^- + k_-^+) + \Delta_\text{S} (t_{12} + t_{21}^*)) \omega_n),
\\
&\Omega f_{12}^{\ua\da}=-(\Delta_\text{B}^3 + \Delta_\text{S} (-k_-^+ t_{21} + k_+^- t_{21}^* - i (t_{11} - t_{22}^*) \omega_n) +
 \Delta_\text{B} (\Delta_\text{S}^2 + k_-^+ k_+^- + t_{21} t_{21}^* - t_{11} t_{22}^* + \omega_n^2)),
\\
&\Omega f_{11}^{\da\ua}=\Delta_\text{B}^2 \Delta_\text{S} + \Delta_\text{B} (k_+^- t_{12} + k_-^- t_{21} + i (t_{11} + t_{22}) \omega_n) +
 \Delta_\text{S} (\Delta_\text{S}^2 + k_-^- k_+^- - t_{12} t_{21} + T1 t_{22} + \omega_n^2),
\\
&\Omega f_{11}^{\da\da}=0,
\\
&\Omega f_{12}^{\da\ua}=\Delta_\text{B}^3 + \Delta_\text{S} (-k_+^+ t_{12} + k_-^- t_{12}^* + i (t_{11}^* - t_{22}) \omega_n) +
 \Delta_\text{B} (\Delta_\text{S}^2 + k_-^- k_+^+ + t_{12} t_{12}^* - t_{11}^* t_{22} + \omega_n^2),
\\
& \Omega f_{12}^{\da\da}=\Delta_\text{S} k_+^+ t_{11} + \Delta_\text{S} k_+^- t_{11}^* - \Delta_\text{B} t_{11} t_{12}^* + \Delta_\text{B} t_{11}^* t_{21} +
 i (\Delta_\text{B} (-k_+^- + k_+^+) + \Delta_\text{S} (t_{12}^* + t_{21})) \omega_n,\\
   &\Omega=\Delta_\text{B}^4 + \Delta_\text{S}^4 + k_-^- k_-^+ k_+^- k_+^+ - k_-^- k_-^+ t_{12}^* t_{21} - k_+^- k_+^+ t_{12} t_{21}^* +
 t_{12} t_{12}^* t_{21} t_{21}^* +
 \Delta_\text{B} \Delta_\text{S} ((k_-^- - k_-^+) (t_{12}^* + t_{21}) + (k_+^- - k_+^+) (t_{12} + t_{21}^*)) - \nonumber \\&
 k_-^+ k_+^- t_{11}^* t_{22} - t_{11} t_{12}^* t_{21}^* t_{22} - k_-^- k_+^+ t_{11} t_{22}^* - t_{11}^* t_{12} t_{21} t_{22}^* +
 t_{11} t_{11}^* t_{22} t_{22}^* -
 i (k_+^- t_{11}^* t_{12} + k_-^- t_{11} t_{12}^* +   k_-^+ t_{11}^* t_{21} + k_+^+ t_{11} t_{21}^* +  k_-^+ t_{12}^* t_{22} + \nonumber \\&
    k_+^- t_{21}^* t_{22} + k_+^+ t_{12} t_{22}^* + k_-^- t_{21} t_{22}^*) \omega_n + (k_-^- k_+^- +
    k_-^+ k_+^+ + t_{11} t_{11}^* + t_{12} t_{12}^* + t_{21} t_{21}^* +
    t_{22} t_{22}^*) \omega_n^2 + \omega_n^4 +
 \Delta_\text{B}^2 (2 \Delta_\text{S}^2 + k_-^+ k_+^- + k_-^- k_+^+ + \nonumber \\& t_{12} t_{12}^* + t_{21} t_{21}^* - t_{11}^* t_{22} -
    t_{11} t_{22}^* + 2 \omega_n^2) +
 \Delta_\text{S}^2 (k_-^- k_+^- + k_-^+ k_+^+ - t_{12} t_{21} - t_{12}^* t_{21}^* + t_{11} t_{22} + t_{11}^* t_{22}^* +
    2 \omega_n^2).
\end{eqnarray}
\end{subequations}
Here we have defined $k_\pm^+=k\cos (\theta_\text{k}+\theta/2)\pm ik\sin (\theta_\text{k}+\theta/2)$ and $k_\pm^-=k\cos (\theta_\text{k}-\theta/2)\pm ik\sin (\theta_\text{k}-\theta/2)$ according to the main text for a twisted BLG by an amount $\theta$.

\twocolumngrid

\end{document}